\documentclass[useAMS,usegraphicx,usenatbib]{mn2e}
\usepackage{aas_macros}
\usepackage[a4paper,centering, totalwidth=520pt, totalheight=700pt]{geometry}
\usepackage[fleqn]{amsmath}
\usepackage{graphicx}
\usepackage{amssymb}
\usepackage{amsmath}
\usepackage{enumerate}
\usepackage{color}
\usepackage{hhline}

\usepackage{fixltx2e}[1999/12/01]

\usepackage[hyperindex]{hyperref}
\hypersetup{
breaklinks = {false},
colorlinks = {false},
linkcolor={black},
pdfpagemode = {None}, 
pdfborder = {0 0 1},
pdftitle = {Photometric redshift errors, galaxy clustering, and BAO},
pdfsubject = {Large-scale structure of the Universe},
pdfauthor = {Chaves-Montero, J.},
pdfkeywords = {Large-scale structure of the Universe}
}

\newcommand{\erf}{\ensuremath{\mathrm{Erf}}}

\newcommand{\Mpc}{\ensuremath{\mathrm{Mpc}}}
\newcommand{\Gpc}{\ensuremath{\mathrm{Gpc}}}

\newcommand{\vk}{\mathbf{k}}

\newcommand{\mufa}{\ensuremath{\langle {\cal F}^2\rangle_{\hat{\mathbf{k}}}}}
\newcommand{\mufb}{\ensuremath{\langle \mu^2\,{\cal F}^2\rangle_{\hat{\mathbf{k}}}}}
\newcommand{\mufc}{\ensuremath{\langle \mu^4\,{\cal F}^2\rangle_{\hat{\mathbf{k}}}}}

\newcommand{\f}{\ensuremath{{\cal F}}}
\newcommand{\ff}{\ensuremath{{\cal F}(k,\mu)}}

\newcommand{\Msun}{ \ensuremath{\rm M_{\odot}} }

\newcommand{\kms}{\ensuremath{\mathrm{km\;s}^{-1}}}

\title[Photo-$z$ errors, galaxy clustering, and BAO]
{The effect of photometric redshift uncertainties on galaxy clustering and
baryonic acoustic oscillations}
\begin{document}

\author[Chaves-Montero et al.]{
\parbox[h]{\textwidth}
{Jon\'as Chaves-Montero$^1$\textsuperscript{\thanks{\href{mailto:jchavesmontero@anl.gov}{jchavesmontero@anl.gov}}},
Ra\'ul E. Angulo$^2$, \& 
Carlos Hern\'andez-Monteagudo$^2$}
\vspace*{6pt}
\\$^1$ High Energy Physics Division, Argonne National Laboratory, Lemont,
IL 60439, USA.
\\$^2$ Centro de Estudios de F\'isica del Cosmos de Arag\'on, Plaza San Juan
1, Planta-3, 44001, Teruel, Spain.
}

\maketitle

\date{\today}

\begin{abstract}
In the upcoming era of high-precision galaxy surveys, it becomes necessary to
understand the impact of redshift uncertainties on cosmological observables. In
this paper we explore the effect of sub-percent photometric redshift errors
(photo-$z$ errors) on galaxy clustering and baryonic acoustic oscillations
(BAO). Using analytic expressions and results from $1\,000$ $N$-body
simulations, we show how photo-$z$ errors modify the amplitude of moments of the
2D power spectrum, their variances, the amplitude of BAO, and the cosmological
information in them. We find that: a) photo-$z$ errors suppress the clustering
on small scales, increasing the relative importance of shot noise, and thus
reducing the interval of scales available for BAO analyses; b) photo-$z$ errors
decrease the smearing of BAO due to non-linear redshift-space distortions (RSD)
by giving less weight to line-of-sight modes; and c) photo-$z$ errors (and
small-scale RSD) induce a scale dependence on the information encoded in the BAO
scale, and that reduces the constraining power on the Hubble parameter. Using
these findings, we propose a template that extracts unbiased cosmological
information from samples with photo-$z$ errors with respect to cases without them.
Finally, we provide analytic expressions to forecast the precision in measuring
the BAO scale, showing that spectro-photometric surveys will measure the
expansion history of the Universe with a precision competitive to that of
spectroscopic surveys.
\end{abstract}



\section{Introduction}

A new generation of wide-field cosmological galaxy surveys will soon map the
spatial distribution of hundreds of millions of galaxies over a wide range of
redshifts. With these, it will be possible to characterise the expansion history
of the Universe and the growth of structures with exquisite precision. Moreover,
these measurements will set strong constraints on the contributors to the total
energy density as a function of redshift, the law of gravity on large scales,
and perhaps will offer hints to explain the accelerated expansion of the
Universe \citep[see ][for a review]{Weinberg2013}.

Some of these future galaxy surveys will employ spectrographs, which will
deliver precise estimates for galaxy redshifts, e.g. Dark Energy Spectroscopic
Instrument \citep[DESI,] [] {DESI16}, WEAVE \citep[] [] {weave14}, Euclid
\citep{euclid11}, and 4-metre Multi-Object Spectroscopic Telescope
\citep[4MOST,] []{4most11}. Other surveys, instead, will rely on either linear
variable filters or sets of narrow-band filters, e.g. the Physics of the
Accelerating Universe Survey \citep[PAUS,][]{pau14}, the Javalambre Physics of
the accelerating universe Astrophysical Survey \citep[J-PAS,][]{benitez14}, and
the Spectro-Photometer for the History of the Universe, Epoch of Reionization,
and Ices Explorer \citep[SPHEREx,] [] {dore14, dore16}. The advantages of the
latter class are: higher surveying speeds, spectral data for every region of the
sky, and a larger number of characterised objects. On the other hand, this
approach adds non-negligible uncertainties to the measured redshifts, which
propagates to the observed galaxy density field. In order to fully exploit the
potential of this class of surveys, the effect of noisy redshift estimators on
galaxy clustering has to be carefully studied.

The impact of photometric redshift errors (photo-$z$ errors) on galaxy
clustering and baryonic acoustic oscillations (BAO) has been investigated by
several authors \citep[e.g.,][] {SeoEisenstein2003, GlazebrookBlake2005,
BlakeBriddle2005, Dolney2006, SeoEisenstein2007, Cai2009, Benitez2009,
Sereno2015, Ross17}. These studies have concluded that, in configuration space,
adding photo-$z$ errors to galaxy redshifts can be regarded as a smoothing of
the galaxy density field along the line-of-sight (LOS). The analogous effect in
Fourier space is a suppression of LOS $k$-modes. Despite this, these authors
demonstrated that BAO can still be detected and used to measure the expansion
history of the Universe through the Hubble parameter $H(z)$ and the angular
diameter distance $D_A(z)$ \citep[e.g.,][]{SeoEisenstein2003}. For instance, for
the same number density, the uncertainty on the measured acoustic scale only
doubles for photo-$z$ errors of $0.3\,\%$ with respect to a spectroscopic case
\citep{Cai2009}. 

In this paper we extend previous studies by developing a complete framework to
extract cosmological information from BAO analyses under the presence of
photo-$z$ errors. In the first half of the manuscript, we explore the problem
analytically. We provide expressions for power spectrum moments (related, but
not identical, to Legendre multipoles) and their variances; we study the
respective signal-to-noise ratios (${\rm SNR}$); and we quantify the
cosmological information encoded in BAO, all this as a function on the
photo-$z$ errors and number density of a given sample. In addition, we provide a
formula that forecasts the precision with which $H(z)$ and $D_A(z)$ can be
measured, as function number density, linear bias, and typical photo-$z$
uncertainty. In the second half of the paper, we use these results to build an
unbiased estimator of the BAO scale from data with photo-$z$ errors. We test the
method by applying it to multiple samples drawn from a set of $1\,000$
cosmological $N$-body simulations.

Our paper is organised as follows: in \S\ref{sec:1o} we describe our cosmological
simulations and how we compute clustering statistics and mimic photo-$z$ errors. 
In \S\ref{sec:2o} we derive analytic expressions for the impact of photo-$z$ errors
on power spectrum moments and their variances, which we compare with the
results from our set of simulations. Then, in \S\ref{sec:3o}, we model how
photo-$z$ errors modify the suppression of BAO and the cosmological information
that they encode. In \S\ref{sec:4o} we build an unbiased model for extracting
the contribution of BAO to power spectrum moments, and in \S\ref{sec:5o} we
apply it to simulated samples with different number densities and photo-$z$
errors. In \S\ref{sec:6o} we make forecasts for the precision with which
cosmological parameters can be measured from galaxy surveys with sub-percent
photo-$z$ errors and in \S\ref{sec:7o} we summarise our most important results.

\section{Numerical Methods} \label{sec:1o}

In this section we present the numerical simulations that we analyse, we explain
how we measure power spectrum moments and their respective variances, and we
describe how we mimic photo-$z$ errors in the simulations.

\subsection{Numerical Simulations}

Typically, numerical simulations poorly sample the modes where BAO are located
\citep[but see][]{AnguloPontzen2016}. Additionally, owing to periodic boundary
conditions, simulations do not consider the coupling with modes larger than the
box size. To avoid these complications and obtain accurate results, it is
necessary to consider ensembles of $N$-body simulations of volumes in excess of
$1\,h^{-3}\Gpc^{3}$ when analysing the BAO feature \citep{Angulo2008}. 

In this work we have carried out an ensemble of $1\,000$ $N$-body simulations,
where each of them evolved $1\,024^3$ dark matter (DM) particles of mass
$1.7\times 10^{12}\,h^{-1}\,\Msun$ in a cubic box of $3\,h^{-1}\Gpc$ on a side
from different initial conditions. This suite has an aggregated volume of
$27\,000\,h^{-3}\Gpc^3$, which will allow us to accurately resolve the BAO
feature.

For computational efficiency, we carried out these simulations using the
Comoving Lagrangian Acceleration (COLA) method \citep{Tassev2013}. This
algorithm is able to recover the real-space power spectrum of a full $N$-body
simulation to within $2\,\%$ for $k<0.3\,h\,\Mpc^{-1}$ at a fraction of its
computational cost \citep{Howlett2015}. Moreover, COLA reproduces the
redshift-space power spectrum monopole and quadrupole of HOD galaxies for
$k < 0.2\,h\,\Mpc^{-1}$ \citep{Koda2015}.

We adopt Gaussian initial conditions created using 2nd-order Lagrangian
Perturbation theory. Gravitational forces were computed using a Particle-Mesh
algorithm with a Fourier grid of $1\,024^3$ points, and particles were evolved
from $z=9$ down to $z=1$ in 10 time steps. The cosmological parameters adopted
were $\Omega_m = 0.25$, $\Omega_\Lambda = 0.75$, $\Omega_b = 0.045$, $n_s=1$,
$H_0 = 73 \,\kms\,\Mpc^{-1}$, and $\sigma_8=0.9$. Each simulation took $3$ CPU
hours to complete.

The COLA ensemble will allow us to investigate the impact of photo-$z$ errors on
power spectrum moments, their variances, and BAO. We will only explore the $z=1$
outputs, which is motivated by the target redshift of future surveys, but our
results can be readily generalised to any redshift and for a full lightcone. In
addition, we will only focus on dark matter statistics, since the typical number
density of halos that can be resolved with at least $100$ particles in our
simulations, $n=1.17\times 10^{-6}\,h^3\,\Mpc^{-3}$, is too low for clustering
studies (thus BAO scales are dominated by shot noise). Nevertheless,
extrapolating our results to biased tracers can be done by considering the
adequate number density and fluctuations amplitude.


\subsection{Power spectrum and covariance measurements}
\label{sec:1b}

Throughout this paper we study the matter density field in Fourier space using
its power spectrum, $P(\mathbf{k})$, defined by

\begin{equation}
\left<\hat{\delta}(\mathbf{k})\hat{\delta}(\mathbf{k}')\right>=
(2\pi)^3 \delta_{\rm D}(\mathbf{k}-\mathbf{k}')\,P(\mathbf{k}),
\end{equation}

\noindent where $\left<\;\right>$ indicates an ensemble average, $\delta_{\rm
D}()$ is the Dirac delta function, and $\hat{\delta}(\mathbf{k})$ is the
Fourier transform of the density contrast field, $\delta(\mathbf{x})$.
Operationally, we compute $P(\mathbf{k})$ by gridding the DM particles of each
simulation onto a $1\,024^3$ cubic lattice using a cloud-in-cell (CIC) scheme
\citep{Hockney81}. Then, we Fast Fourier Transform this field and correct for
the assignment scheme by dividing by the Fourier transform of the CIC window
function. We expect this to provide power spectrum estimates accurate to within
$0.1\,\%$ up to $k=0.4\,h\,\Mpc^{-1}$ \citep{Sefusatti16}.

Within the plane-parallel approximation, the 3D power spectrum has an azimuthal
symmetry and it can be decomposed in terms of the following moments:

\begin{equation}
P_{\ell}(k) = \frac{1}{2}\int_{-1}^{1} \text{d}\mu\,\mu^\ell P(k,\mu),
\end{equation}

\noindent where $k\equiv|\mathbf{k}|$ is the modulus of the wave-vector
$\mathbf{k}$, and $\mu=\frac{\mathbf{k}\cdot \hat{{\bf k}_z}}{k}$. 

In the Kaiser approximation, the full 2D power spectrum can be written using
combinations of $\ell = 0$, 2, and 4 moments only. Even when considering
nonlinearities, almost all cosmological information is encoded in these
three moments, which is why we will only consider these in our subsequent 
analysis.

Note that we chose to adopt moments of the power spectrum, rather than the more
common Legendre multipoles, for simplicity in the analytic expressions and in
the numerical analysis we will present later. In practice, this choice is unimportant
as any Legendre multipole of order $\ell'$ can be written as a linear combination of
moments of order $\ell \leq \ell'$. 

To minimise the impact of a discrete sampling of wavemodes, specially on large
scales, we compute $P_{\ell}(k)$ using a $o$-point Gauss-Legendre quadrature
algorithm \citep{Abramowitz72,Szapudi01, Kashlinsky01}:

\begin{equation}
\label{eq:compmono}
P_{\ell}(k) \simeq \frac{1}{2}\sum_{i}^o w_i\,y_i^\ell \hat{P}(k,y_i),
\end{equation}

\noindent where $w_i=\frac{2}{(1-y_i^2)[\mathcal{L}'_o]^2}$ is a Gauss-Legendre
weight, $\mathcal{L}_o$ is the $o$-th order Legendre polynomial, and $y_{j}$ is
the $j$-root of the Legendre polynomial of order $o$. We estimate $\hat{P}(k,y)$
from neighbouring measurements:

\begin{equation}
\label{Pmono_est}
\hat{P}(k,y) = \frac{\sum_{\vk_i} \sum_j^o p(\mu_i,y_j) P(\vk_i)^2}
{\sum_{\vk_i} \sum_j^{o} p(\mu_i,y_j)},
\end{equation}

\noindent where $\mu_i$ is the direct cosine of $\mathbf{k}_i$, and the sum
$\sum_{\vk_i}$ runs over the $N_k$ wave-vectors $\vk_i$ that lie within a bin in
$k$, which we define to be equally spaced in $\Delta k = (2\pi/L_{\rm
box})h\,\Mpc^{-1}$. The terms $p(\mu_i, y_j)=e^{-0.5(\mu_i-y_j)^2/(\Delta
p_j)^2}$ are arbitrary weights, which we set by choosing $\Delta p_j$ to be
twice the distance between two consecutive roots, $\Delta p_j = 2(y_{j+1}-y_j)$.
We have checked that $o=16$ is enough for accurate results. We note that on very
large scales this algorithm introduces a small covariance between different
$\mu$-bins, owing to the limited number of available $k$-modes.

Finally, we apply a correction to every moment to remove at first
order the contribution of shot noise, $P_\ell \rightarrow P_\ell -
n^{-1}/(\ell+1)$, where $n$ is the average number density of objects
considered.

An ensemble of $M$ measurements can be used to compute the respective covariance
matrix:

\begin{equation}
\mathbf{C}_{\ell}(k_i,k_j) = \frac{1}{M-1}
\sum_{m=1}^M [P_{\ell}^m(k_i)-
\bar{P_{\ell}}(k_i)][P_{\ell}^m(k_j)-\bar{P_{\ell}}(k_j)],
\end{equation}

\noindent where $P_{\ell}^m$ is the $\ell$-th moment of the $m$-th simulation
and $\bar{P_{\ell}}$ its average estimated from $M$ simulations. In
\S\ref{sec:5o} we will extract the BAO scale jointly from $P_0$, $P_2$, and
$P_4$. In that case, we substitute the vector $P_\ell^m$ by
$\{P_0^m,P_2^m,P_4^m\}^{\rm T}$.

We calculate the precision matrix, $\mathbf{C}_\ell^{-1}$, by inverting 
the covariance matrix, $\tilde{\mathbf{C}}_{\ell}^{-1}$, using an algorithm
based on a LU factorisation. The expected value of this matrix is biased 
when computed from a finite number of realisations. We correct for this as follows:

\begin{equation}
\mathbf{C}_{\ell}^{-1} = 
\frac{M - N_{\rm bins} - 2}{M - 1}\,\tilde{\mathbf{C}}_{\ell}^{-1},
\end{equation}

\noindent where $N_{\rm bins}$ is the number of $k$-bins in $P_\ell$
\citep{Hartlap2007}.

\subsection{Redshift uncertainties}
\label{sec:1c}

We model redshift-space distortions (RSD) and photo-$z$ errors in our
simulations in the flat sky approximation, i.e. we perturb the comoving position
of objects along the $\hat{{\bf k}_z}$ direction:

\begin{equation}
s_z = r_z + (1+z_{\rm box})\frac{{v}_z}{H(z_{\rm box})} + \delta(r_z)
\end{equation}

\noindent where $s_z$ and $r_z$ are the perturbed and unperturbed comoving
positions, respectively, $v_z$ is the physical peculiar velocity along the
$z$-axis in $\kms$, $H(z_{\rm box})$ the Hubble parameter at the redshift of the
simulation box in $\kms \Mpc^{-1}$, and $\delta(r_z)$ a random variable that
account for photo-$z$ errors and whose PDF given by $\Pr[\delta(r_z)]$.

Unless stated otherwise, in what follows we assume that $\Pr[\delta(r_z)]$ is a
Gaussian distribution with zero mean and standard deviation $\sigma =
\sigma_z(1+z_{\rm box})\,c\,H^{-1}(z_{\rm box})$ where $\sigma_z$ indicates the
precision in redshift.



\section{Effect of photometric redshift errors on power spectrum moments}
\label{sec:2o}

In this section we derive analytic expressions for the impact of photo-$z$
errors on power spectrum moments and their variances. In all cases, we compare
these predictions with numerical results obtained from the COLA ensemble.


\subsection{Power spectrum moments} \label{sec:2a}

\subsubsection{General expressions} \label{sec:2aa}

Let us consider a set of galaxies with a real-space overdensity
$\hat{\delta}_r(\mathbf{k})$ discretely sampling a field of covariance
$P(\mathbf{k})$, and whose redshifts are measured through a noisy but unbiased
estimator. The observed redshifts are thus $z + \delta z$, where $\delta z$ is
the photo-$z$ error. Assuming that the PDF of the photo-$z$ errors, $\Pr[\delta
r_z]$, is identical for every galaxy, the redshift-space overdensity field
within the Gaussian dispersion model \citep{Kaiser1987, Peacock1994} is:

\begin{align}
\hat{\delta}_z(k,\mu) =& \hat{\delta}_r(k) \ff, \\
\ff \equiv&  (1+\beta \mu^2) \, 
e^{-0.5[k \mu \sigma_v(1+z)c/H(z)]^2} F(k \mu) \label{eq:facrr},
\end{align}

\noindent where $\beta \equiv b^{-1}\,{\rm d\,ln\,}D(a)/{\rm d\,ln\,}a$, $b$ is
the large-scale bias of the sample, $D(a)$ is the linear growth factor for 
dark matter, $a=(1+z)^{-1}$ is the cosmological scale factor, $\sigma_v$ is a
velocity dispersion induced by non-linear dynamics in units of the speed of
light, and $F(k\mu)$ is the Fourier transform of $\Pr[\delta r_z]$. The first
and second term of the RHS of Eq.~\ref{eq:facrr} encode large- and small-scale
RSD generated by the peculiar velocity of the galaxies, respectively, and the
third the effect of photo-$z$ errors. Hereafter, for brevity we will not write
explicitly the dependence of $\ff$ on $\mu$ and $k$.

We have chosen to adopt a Gaussian form for distribution of small-scale
velocities. However, other forms better fit the distributions measured in
cosmological simulations \citep{Scoccimarro04, orsiangulo2017}. Our results can
be easily generalised to those distributions by replacing the exponential term
in Eq.~ \ref{eq:facrr} by the desired velocity distribution function.

From Eqs.~2,8, and 9, we can derive the relation between power spectrum moments in
redshift space, $P_\ell$, and the $l=0$ moment of the real-space power spectrum,
$P^r_0$:

\begin{equation}
P_\ell(k) = \langle \mu^\ell\, \f^2
\rangle_{\hat{\vk}} P^r_0(k), \label{eq:pkmumodel1}
\end{equation}

\begin{equation}
\langle \mu^\ell\, \f^2 \rangle_{\hat{\vk}} = \frac{1}{2}\int_{-1}^{1} 
\text{d}\mu\,\mu^\ell\,\f^2, \label{eq:pkmumodel2}
\end{equation}

\noindent where, here and in the remainder of this paper, the $\langle ...
\rangle_{\hat{\vk}}$ brackets denote an angular average. 

\begin{figure}
\begin{center}
\includegraphics[width=0.42\textwidth]{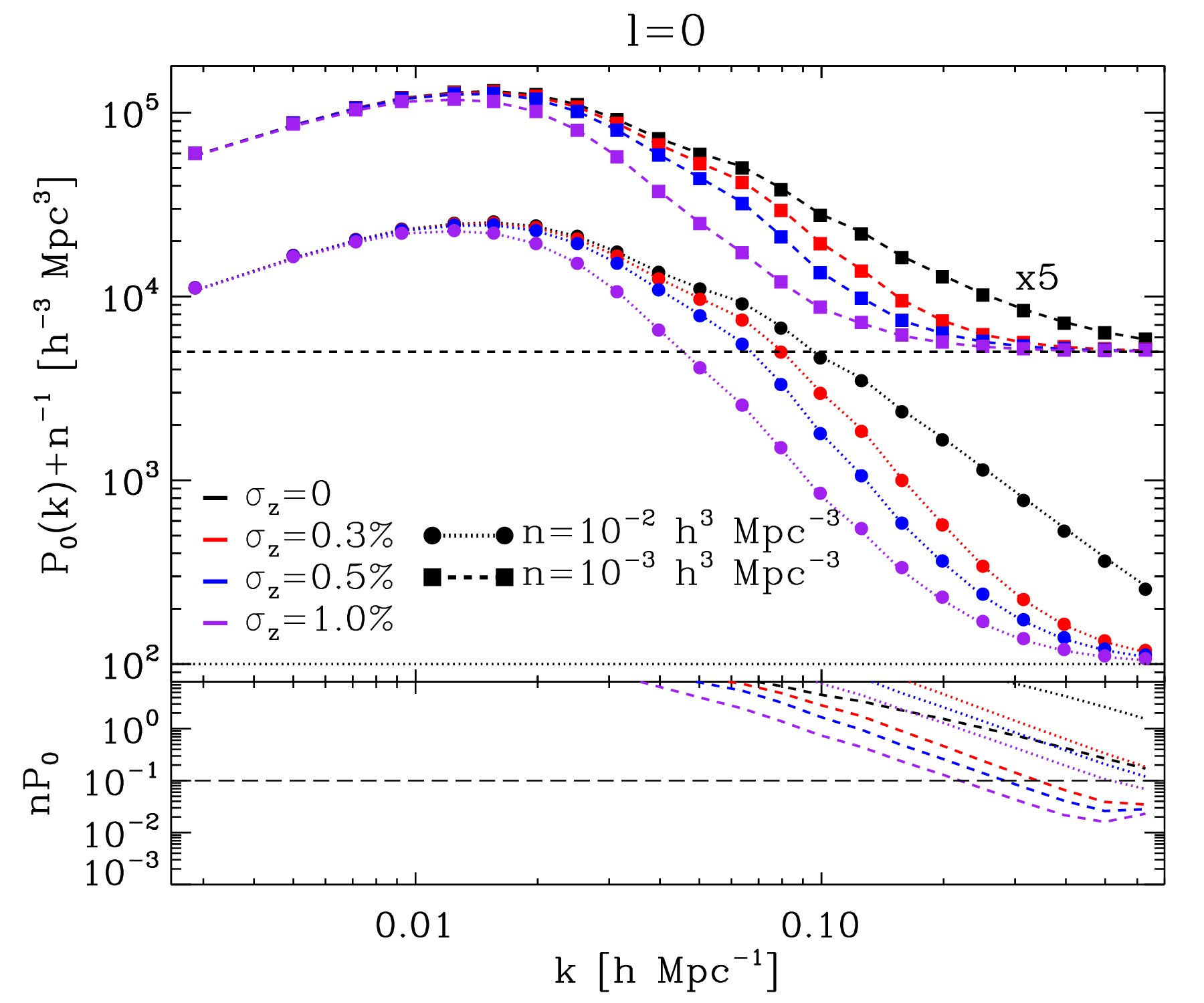}

\includegraphics[width=0.42\textwidth]{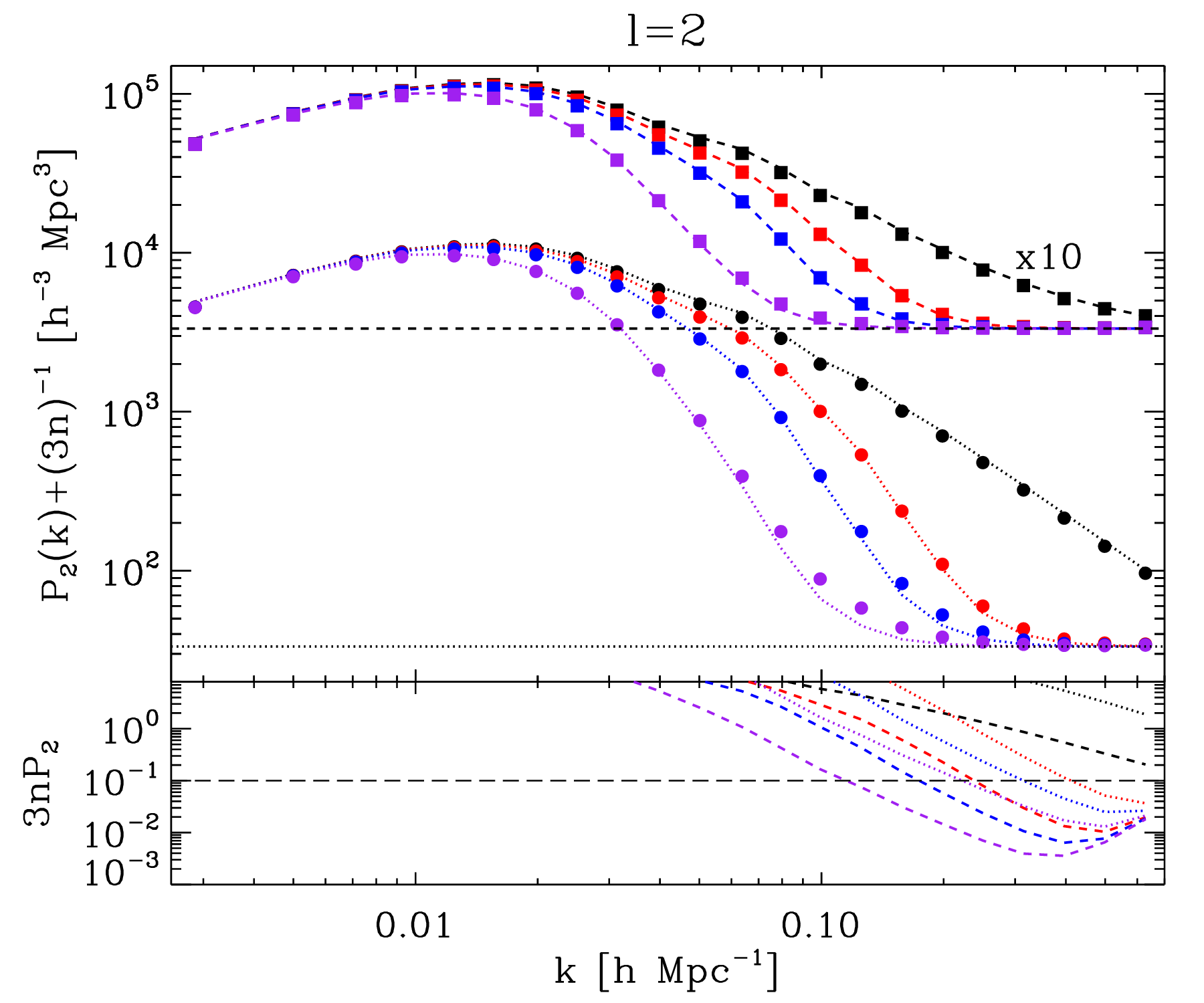}

\includegraphics[width=0.42\textwidth]{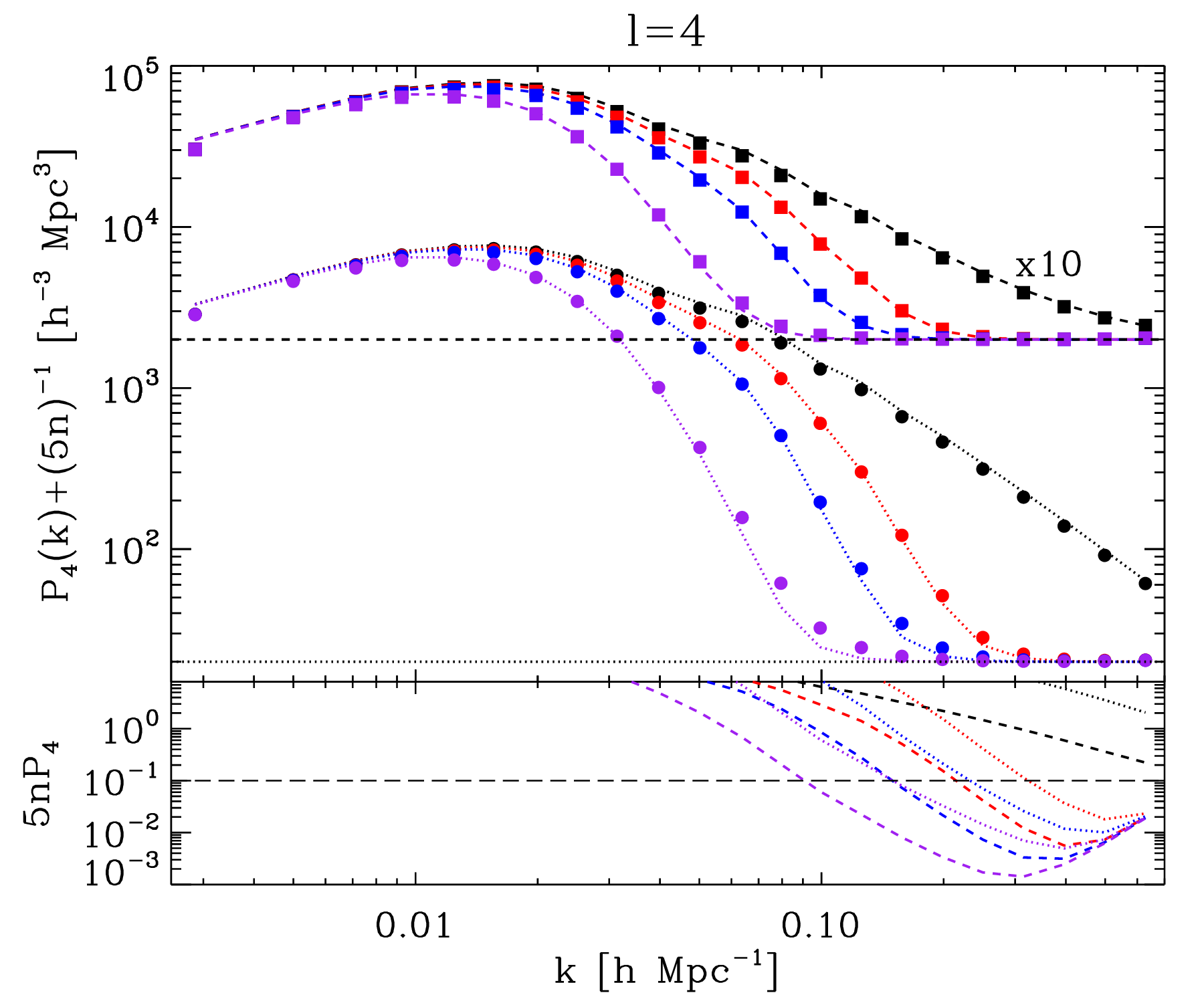}
\end{center}
\caption{
\label{fig:1}
{\bf Top panels:} impact of photo-$z$ errors on $P_0$, $P_2$, and $P_4$. Symbols
show the average results from an ensemble of $1\,000$ $N$-body simulations and
lines our analytic predictions (Eqs.~\ref{eq:pkmumodel1}--\ref{eq:p4zph1}),
which employ as input the average $P^r_0$ from simulations with
$n=0.03\,h^3\,\Mpc^{-3}$ and $\sigma_v=3\times 10^{-4}$. The colour and shape of
lines and symbols indicate the size of Gaussian photo-$z$ errors and the number
density for each sample, respectively, as stated in the legend. Results for
different number densities are vertically displaced for clarity. We employ this
colour-coding in what follows. Horizontal lines indicate the shot noise level.
Our analytic model precisely captures the effect of photo-$z$ errors. {\bf
Bottom panels:} ratio of shot noise corrected power spectrum moments from
simulations and shot noise. On the scales where $(2\ell+1)nP_\ell<0.1$
(indicated by long-dashed lines) the shot noise correction is no longer
accurate.}
\end{figure}

\begin{figure}
\begin{center}
\includegraphics[width=0.37\textwidth]{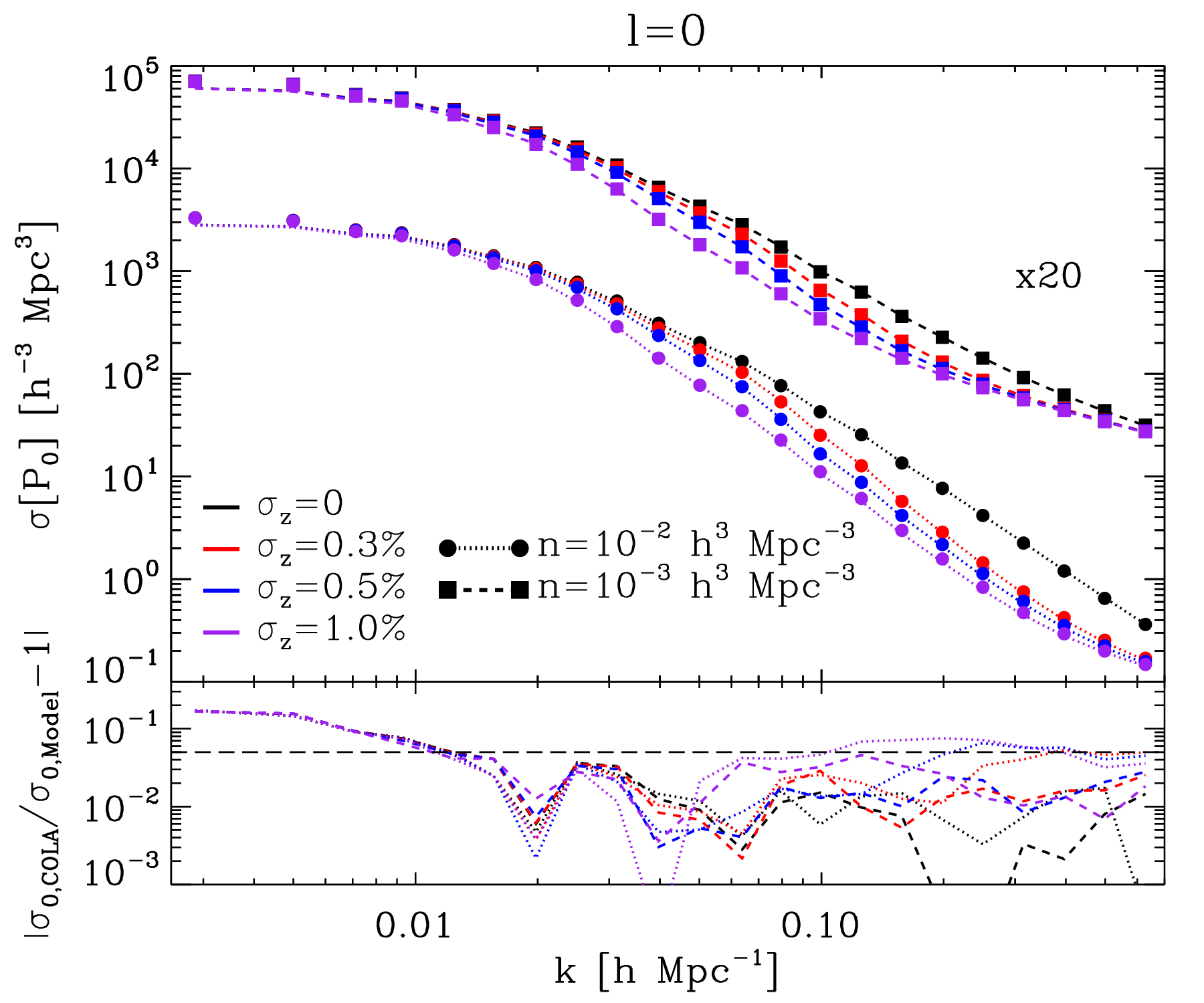}

\includegraphics[width=0.37\textwidth]{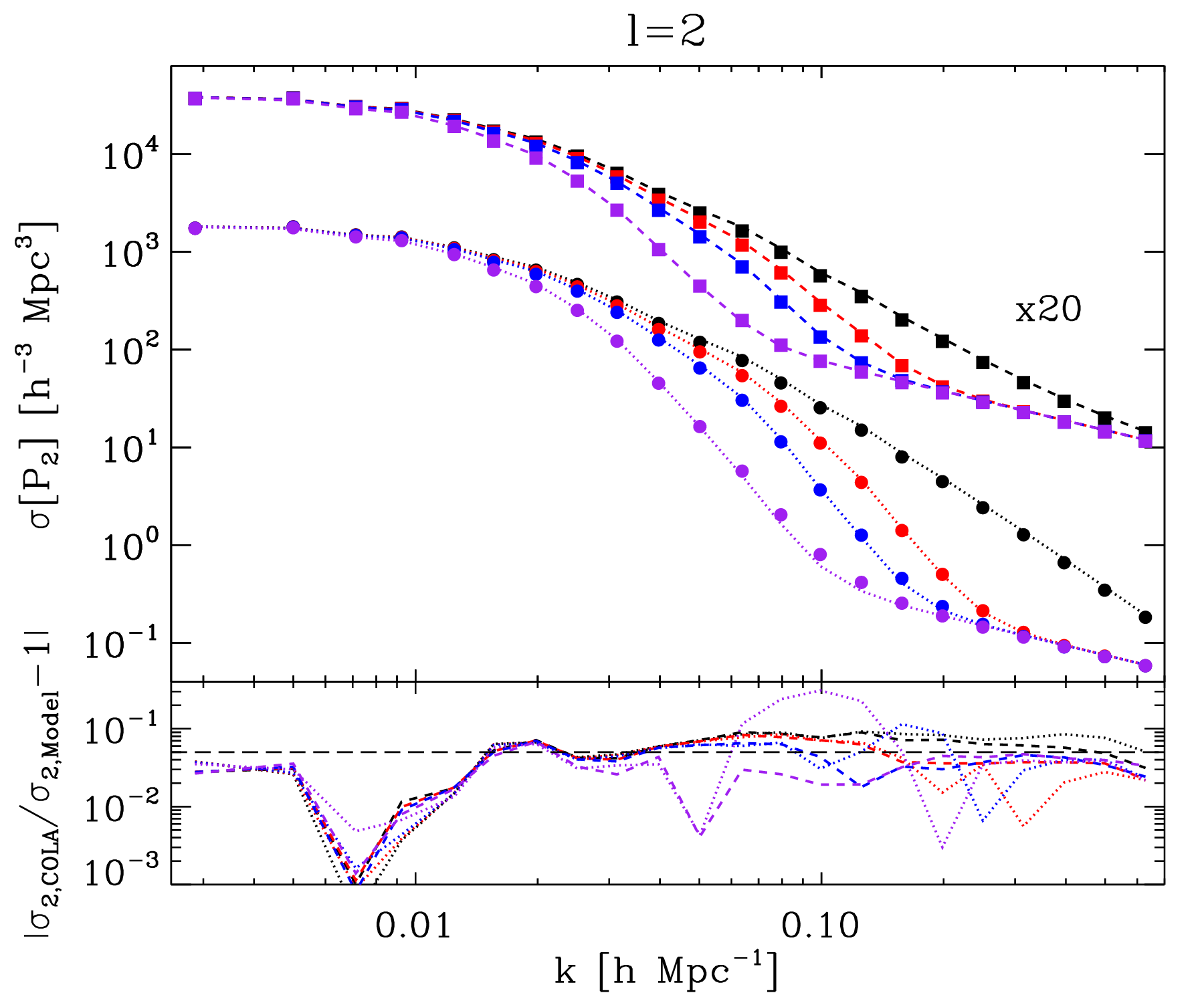}

\includegraphics[width=0.37\textwidth]{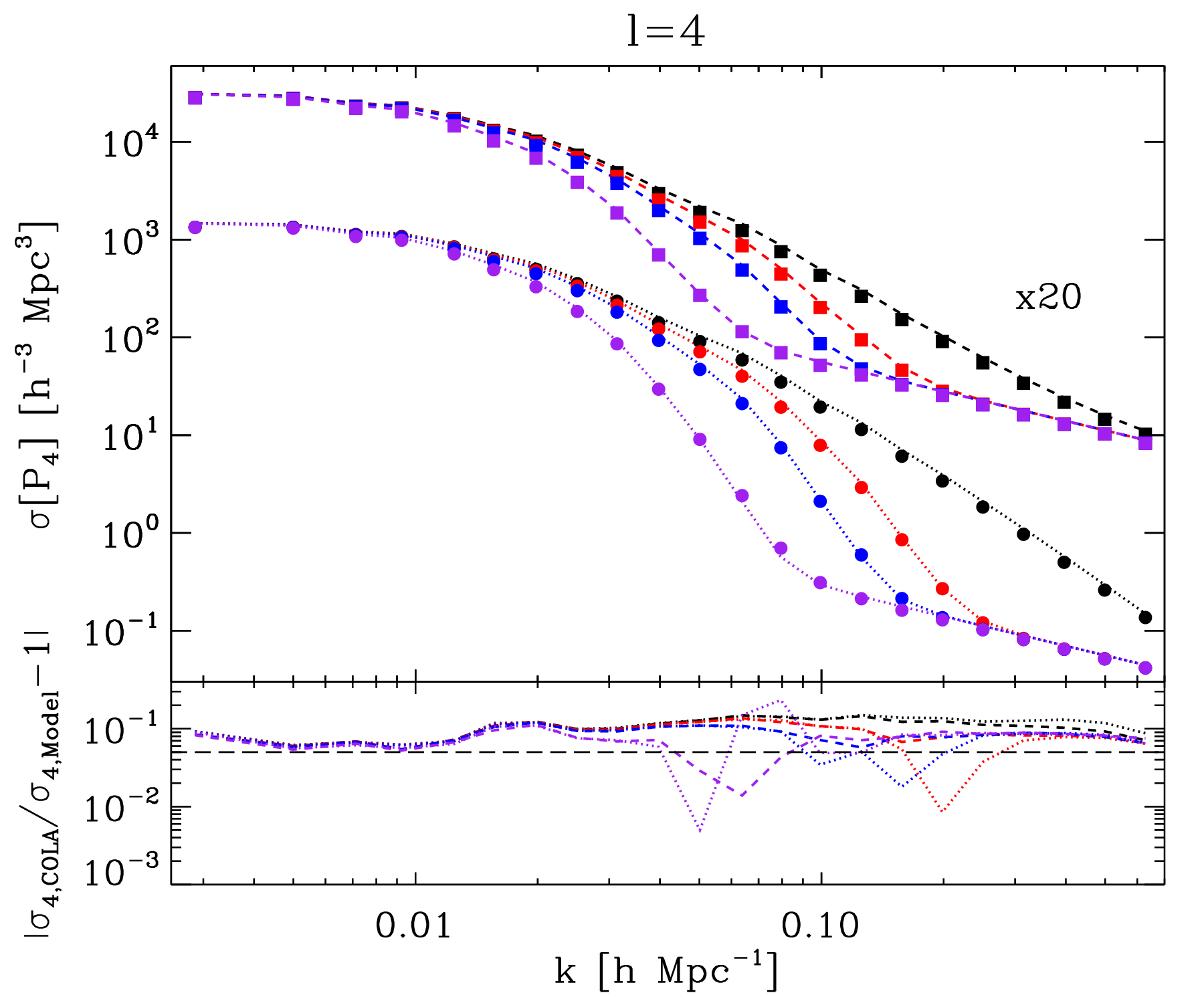}
\end{center}
\caption{
\label{fig:2} The impact of photometric errors on the variance of the moments of
the redshift-space power spectrum. In each panel we show three sets of curves
displaying the results for different number densities. Within each set, colours
indicate different redshift uncertainties, as indicated by the legend. Symbols
indicate the mean values from our ensemble of simulations, whereas dotted lines
indicate the analytic model of Eqs. ~\ref{eq:smodel}. The bottom panel of each
figure shows the fractional difference between our numerical and analytic
results.}
\end{figure}

\subsubsection{The Gaussian case} \label{sec:gaussian_case_p}

For a Gaussian $\Pr[\delta r_z]$, $F(k \mu) = \exp \{-0.5[k\,\mu
\,\sigma_z (1+z) c/H(z)]^2\}$, and the $\ell=0,\,2,\,{\rm and}\,4$ moments 
can be derived analytically:

\begin{align}
\label{eq:pzph1}
\mufa=&\frac{\sqrt{\pi}}{2}\frac{\erf(x)}{x}
\left(1+\frac{\beta}{x^2}+\frac{3\beta^2}{4x^4}\right) \nonumber \\
&-\frac{\beta e^{-x^2}}{x^2} \left(1+\frac{3\beta}{4x^2}\mathcal{H}_1(x)\right), 
\end{align}

\begin{align}
\label{eq:p2zph1}
\mufb =& \frac{\sqrt{\pi}}{4}\frac{\erf(x)}{x^3}
\left(1 +\frac{3\beta}{x^2}+\frac{15\beta^2}{4x^4}\right) \nonumber \\
&-\frac{e^{-x^2}}{2x^2} \left(1+ \frac{3\beta}{x^2}\mathcal{H}_1(x)
+\frac{15\beta^2}{4x^4}\mathcal{H}_2(x)\right),
\end{align}

\begin{align}
\label{eq:p4zph1}
\mufc =& \frac{3\sqrt{\pi}}{8}\frac{\erf(x)}{x^5}
\left(1 +\frac{5\beta}{x^2}+\frac{35\beta^2}{4x^4}\right) \nonumber \\
&-\frac{3e^{-x^2}}{4x^4} \left(\mathcal{H}_1(x)+ 
\frac{5\beta}{x^2}\mathcal{H}_2(x)
+\frac{35\beta^2}{4x^4}\mathcal{H}_3(x)\right),
\end{align}

\noindent where $\mathcal{H}_n(x)=\sum_{i=0}^n \frac{2^i}{(2i+1)!!}\,x^{2i}$,
$!!$ denotes the double factorial, $x=k\,\sigma_{\rm eff}$, and $\sigma_{\rm
eff} = \sqrt{\sigma_z^2 + \sigma_v^2}\,(1+z)\,c/H(z)$, i.e. the redshift
uncertainties and small-scale peculiar velocities are added in quadrature
\citep{Peacock1994}. Note that these expressions diverge as $x\to0$, and in
general $\langle \mu^n\,\f^m \rangle_{\hat{\vk}}$ expressions are only valid
when $x>3$. To obtain valid expressions when $x<3$, we expand $F(k \mu)$ into a
power series. 

The expressions in real space can be trivially obtained by setting $\beta = 0$
and $\sigma_{\rm eff}=\sigma$, recovering the expression for Eq.~\ref{eq:pzph1}
provided in \citet{Peacock1994}. In that case, we can see that photo-$z$ errors
always generates an apparent anisotropic clustering, even if the underlying galaxy
field is isotropic. In redshift space the effect is more complex as 
photo-$z$ errors couple with RSD parameters. We will explore this next.

\subsubsection{Comparison with numerical simulations}

In the top panels of Fig.~\ref{fig:1} we display the average power
spectrum moments from our analytic expression and from the COLA ensemble. Symbols 
and colours indicate the results for samples with different number densities and
photo-$z$ errors, as stated in the legend. To compute our model, we employ the 
average $P^r_0$ from the COLA ensemble with $n=0.03\,h^3\,\Mpc^{-3}$ and 
$\sigma_v=3\times10^{-4}$, where $\sigma_v$ is obtained by fitting our analytic model
to power spectrum moments of samples without photo-$z$ errors. 

We can see that our numerical and analytical results show good agreement. The
discrepancies, of the order of $10\,\%$, arise from the inaccuracy of our RSD model and 
shot noise subtraction. In the bottom panels  of Fig.~\ref{fig:1} we display 
the ratio of shot noise  corrected moments and their shot noise level. The long dashed 
line indicates when the shot noise level is 10 times greater than the amplitude of
shot noise corrected moments. As we can see, below this line the shot noise 
subtraction is no longer precise.

Overall, we can see that photo-$z$ errors suppress the amplitude of $P_0$, $P_2$, and
$P_4$ for all wavenumbers, specially on small scales. The Poisson noise, however, in
unaltered. Therefore, for a given number density of objects, photo-$z$s reduce the
number of modes and scales useful for cosmological analyses. This is the main effect
for galaxy clustering.


\subsection{Variance of power spectrum moments}
\label{sec:2b}

\subsubsection{General expressions}

The effect of photo-$z$ errors is not only to modify the amplitude of the power
spectrum $P$ but also its Gaussian covariance. Let us first consider the
diagonal elements of the power spectrum covariance matrix:

\begin{equation}
\label{eq:ep1}
\sigma^2[P](k) = \frac{2}{N_k}  \sum_{\vk_i} 
\langle |\hat{\delta}(\vk_i)|^4\rangle - \langle \hat{P}(\vk_i) \rangle^2,
\end{equation}

\noindent where $\langle ... \rangle$ denotes the ensemble average over multiple
realisations/universes. The factor two appears because only half of the modes of
the power spectrum are independent due to the reality of $\delta(\mathbf{x})$.

Assuming that the real and imaginary parts of $\hat{\delta}(\mathbf{k})$ are
Gaussian random variables with zero mean and standard deviation $P/2$, and
combining Eqs.~\ref{eq:pkmumodel1} and \ref{eq:ep1}, we obtain the following
expression for the variance of power spectrum moments in redshift space under
the presence of shot noise, photo-$z$ errors, and small-scale velocities:

\begin{equation}
\label{eq:smodel}
\sigma^2[P_\ell] = \frac{2}{N_k}
\biggl[\langle \mu^{2\ell}\f^4\rangle_{\hat{\mathbf{k}}}
+\frac{2\langle \mu^{2\ell} \f^2\rangle_{\hat{\mathbf{k}}}}{n P^r_0}
 + \frac{1}{(2\ell+1)(n P^r_0)^2}\biggr](P^r_0)^2,
\end{equation}

\noindent where this expression reduces to that provided by \citet{Colombi2009}
for $\ell=0$ in real space without photo-$z$ errors ($\f = 1$). Note that to
compute the diagonal terms of the covariance between two power spectrum moments
$P_\ell$ and $P_{\ell'}$, we only have to substitute $2\ell$ by $\ell+\ell'$ in
the previous equation. 

Note that for a Gaussian $\Pr[\delta r_z]$, all terms in Eq.\ref{eq:smodel} have an
analytic expression, which we provided in \S\ref{sec:gaussian_case_p}. We can
thus analytically compute the variance of any power spectrum moment. 

\subsubsection{Comparison with simulations: diagonal terms}
\label{sec:2bc}

In Fig.~\ref{fig:2} we display the variance of $P_0$, $P_2$, and $P_4$
for the same samples shown in Fig.~\ref{fig:1}. In the bottom panels we
compare our analytic model and the results from our simulations. Overall, our
expressions correctly capture the effect of photo-$z$s -- the agreement is to
within $5\,\%$, $10\,\%$, and $20\,\%$ for $P_0$, $P_2$, and $P_4$, respectively. 
The differences originate from assuming that the matter density field is Gaussian, 
and thus from neglecting in our calculations the contribution of a non-zero
trispectrum.

We see that photo-$z$ errors reduce the variance, especially on intermediate scales
where BAO are located, which is analogous to their effect on the amplitude of power 
spectrum moments. On the other hand, at a fixed scale, the contribution of shot noise
progressively dominates as the order of the moment increases. We can understand this
from \ref{eq:smodel}: since the last term in brackets of Eq.~ does not depend on 
photo-$z$ errors, it will be more important as the redshift uncertainty increases.


\begin{figure}
\begin{center}
\includegraphics[width=0.475\textwidth]{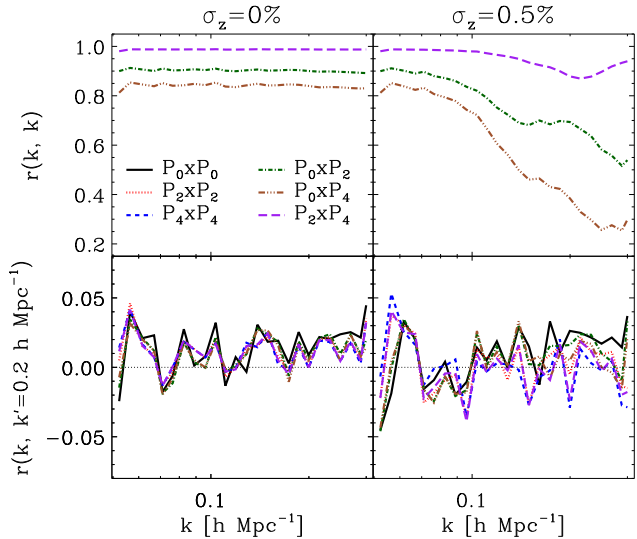}
\end{center}
\caption{
\label{fig:3}
Diagonal and off-diagonal elements (top and bottom panels, respectively) of the
auto-correlation matrices of $P_0$, $P_2$, $P_4$, and their cross-correlations. The
left and right panels show the results for $\sigma_z=0$ and $0.5\,\%$,
respectively, which are computed using $1\,000$ samples from the COLA ensemble
with $n=0.01\,h^3\,\Mpc^{-3}$. For samples with photo-$z$ errors the diagonal
terms of the cross-correlations decrease on small scales (we do not display
those of the auto-correlations because their value is one by definition).
Off-diagonal terms are computed at $k=0.2\,h\Mpc^{-1}$, and their amplitudes are
small independently of the size of photo-$z$ errors.}
\end{figure}


\subsubsection{Comparison with simulations: off-diagonal terms}
\label{sec:2bd}

In Fig.~\ref{fig:3} we display the diagonal and off-diagonal elements of the
auto-correlation matrices of $P_0$, $P_2$, and $P_4$, and their cross-correlations
for $\sigma_z=0$ and $0.5\,\%$. The results shown are computed using $1\,000$
samples from the COLA ensemble with $n=0.01\,h^3\,\Mpc^{-3}$. The top panels
display the cross-correlation between different moments, which is
scale-independent for samples with no errors. However, for samples with
photo-$z$ errors they decrease by increasing the value of $k$. This is the
consequence of photo-$z$ errors modifying the information content of power
spectrum moments, we further analyse this in the following section.

On the bottom panels we display the off-diagonal terms of the correlation
matrices at $k=0.2\,h\Mpc^{-1}$. They are negligible at this scale for samples
with and without photo-$z$ errors. This is because photo-$z$ errors do not
modify the structure of the covariance matrix. In particular, in the
plane-parallel approximation if the real-space power spectrum covariance matrix
is diagonal for samples with no photo-$z$ errors, then so it is in redshift
space with or without photo-$z$ errors. We check that in the whole range of
scales used for BAO analyses in \S\ref{sec:5o} off-diagonal are also smaller
than $5\,\%$, which will justify the employment of Eq.~\ref{eq:intsneff}.
Nonetheless, on even smaller scales the covariance matrix is no longer diagonal
due to non-linearities.


\subsection{Signal-to-noise ratio} \label{sec:2c}

Let us now consider the ${\rm SNR}$ of power spectrum moments, which we define
as the ratio between a given moment and the square root of its variance. We use
this approximation because the covariance matrices of $P_0$, $P_2$, and $P_4$
are mostly diagonal on the scales where BAO are located, as we showed in the
previous section. Photo-$z$ errors decrease the amplitude of power spectrum
moments as well as their variances, and thus the resulting ${\rm SNR}$ will
depend on a balance between both effects.

\subsubsection{Toy model} \label{sec:toy}

In this subsection we introduce a toy model to understand how photo-$z$ errors
modify the ${\rm SNR}$ of $P_0$, where it is straightforward to extend this
model to $\ell>0$. The model is the following:

\begin{equation}
\label{eq:toymono}
{\hat P}_0 = \frac{1}{2} {\hat P}_0^r \biggl[\eta(\mu_1)e^{-k^2 \sigma_{\rm
eff}^2 \mu_{1}^2}+\eta(\mu_2) e^{-k^2 \sigma_{\rm eff}^2 \mu_2^2}\biggr],
\end{equation}

\noindent where the terms in brackets provide the angular contribution at only
two $\mu$-values ($\mu_1$ and $\mu_2$), the symbol ${\hat P}_0^r$ denotes the
measured real-space $\ell=0$ moment, and $\eta(\mu)$ describes the contribution
of large-scale RSD in a $\mu$-bin. We will assume that $\mu_1 < \mu_2$, and thus
$\eta(\mu_1) < \eta(\mu_2)$ since on linear scales $\eta (\mu)$ is a
monotonically increasing function of $\mu$.

For an ensemble average over a given $k$-bin we have that the ${\rm SNR}$ per
radial $k$-interval reads:

\begin{equation}
\label{eq:s2n_toym}
{\rm SNR} =
\frac{1+\eta_{21}\exp(-k^2\,\sigma_{\rm eff}^2\,\Delta \mu^2)}
{\sqrt{1+\eta_{21}^2\exp(-2\,k^2\,\sigma_{\rm eff}^2\,\Delta \mu^2)}},
\end{equation}

\noindent with $\Delta \mu^2 = \mu_2^2- \mu_1^2$ and
$\eta_{21}=\eta(\mu_2)/\eta(\mu_1)$. From this expression, we shall consider
three different cases:

\begin{itemize} 

\item No photo-$z$ errors nor small-scale RSD, $k\,\sigma_{\rm eff} = 0$. In
this case 

\begin{equation}
\label{eq:s2n_case1}
{\rm SNR} =  \frac{1+\eta_{21}}  {\sqrt{1+\eta_{21}^2 }},
\end{equation}

\noindent where the ${\rm SNR}$ is always below $\sqrt{2}$, which is the value
corresponding to real space.

\item Very large photo-$z$ errors, $k\,\sigma_{\rm eff} \to \infty$. In this
limit, the value of the ${\rm SNR}$ is 1. It is smaller than in the first case
because all information along $k$-modes parallel to the LOS is lost.

\item Small photo-$z$ errors, $k\,\sigma_{\rm eff} \rightarrow 0$. In this case,
to first order in $(k\,\sigma_{\rm eff})^2$ we obtain

\begin{equation}
\label{eq:s2n_case3}
{\rm SNR} = \frac{1+\eta_{21}}{\sqrt{1+\eta_{21}^2}}
+\frac{\Delta \mu^2\, \eta_{21}\,(\eta_{21}-1)}
{(1+\eta_{21}^2)^{3/2}}\,(k\,\sigma_{\rm eff})^2
+{\cal O}\bigl[(k\,\sigma_{\rm eff})^4\bigr],
\end{equation}

\noindent where in this limit the ${\rm SNR}$ increases with respect to the case
without photo-$z$ errors nor small-scale RSD as $\eta_{21}>1$. This behaviour
must thus yield a local maximum in the ${\rm SNR}$, since for larger
$k\,\sigma_{\rm eff}$ values we must recover the second case. This reflects that
in this limit photo-$z$ errors affect more the standard deviation of $P_0$ than
its amplitude, and thus they slightly increase the ${\rm SNR}$.

\end{itemize}

\begin{figure}
\begin{center}
\includegraphics[width=0.37\textwidth]{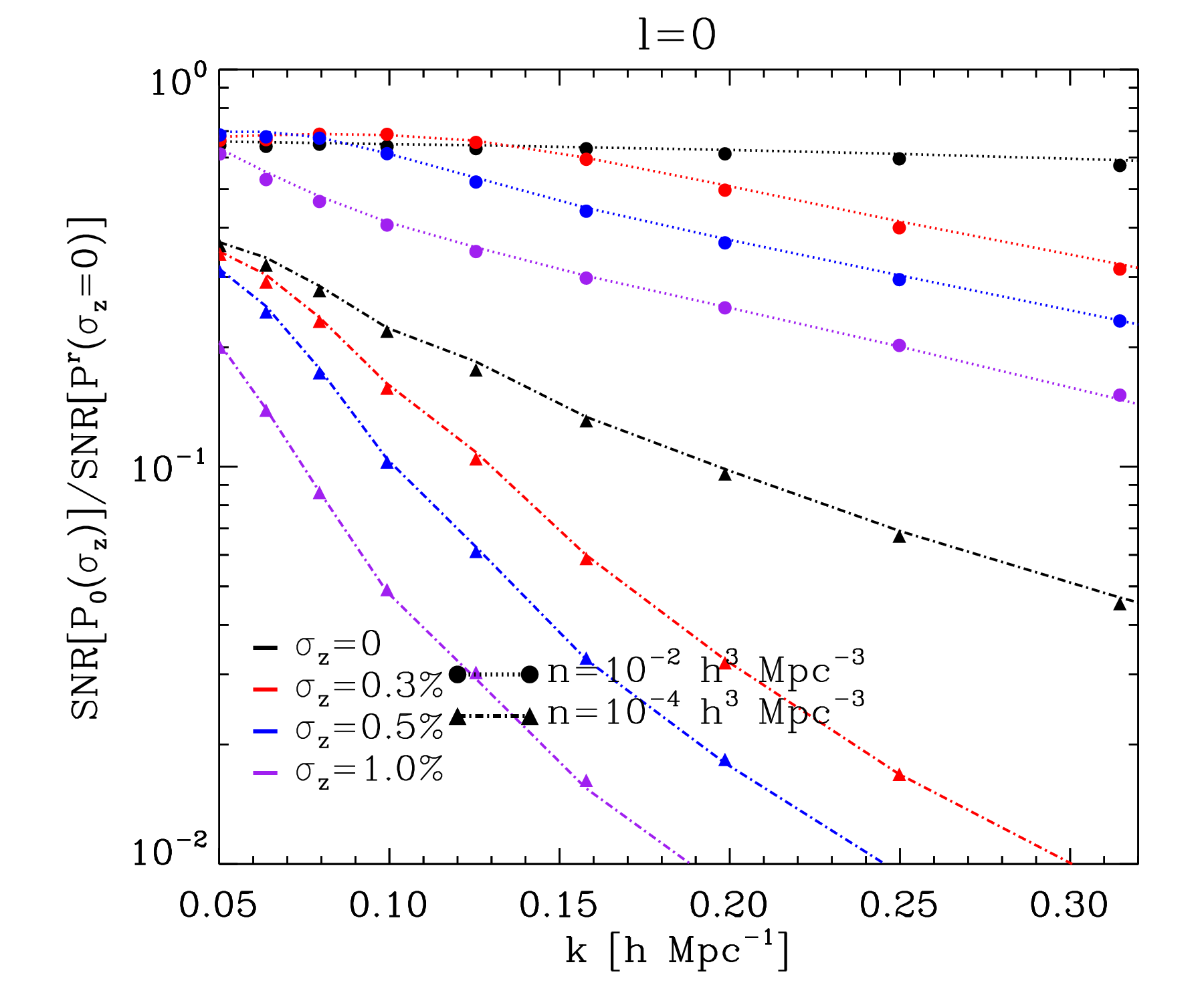}

\includegraphics[width=0.37\textwidth]{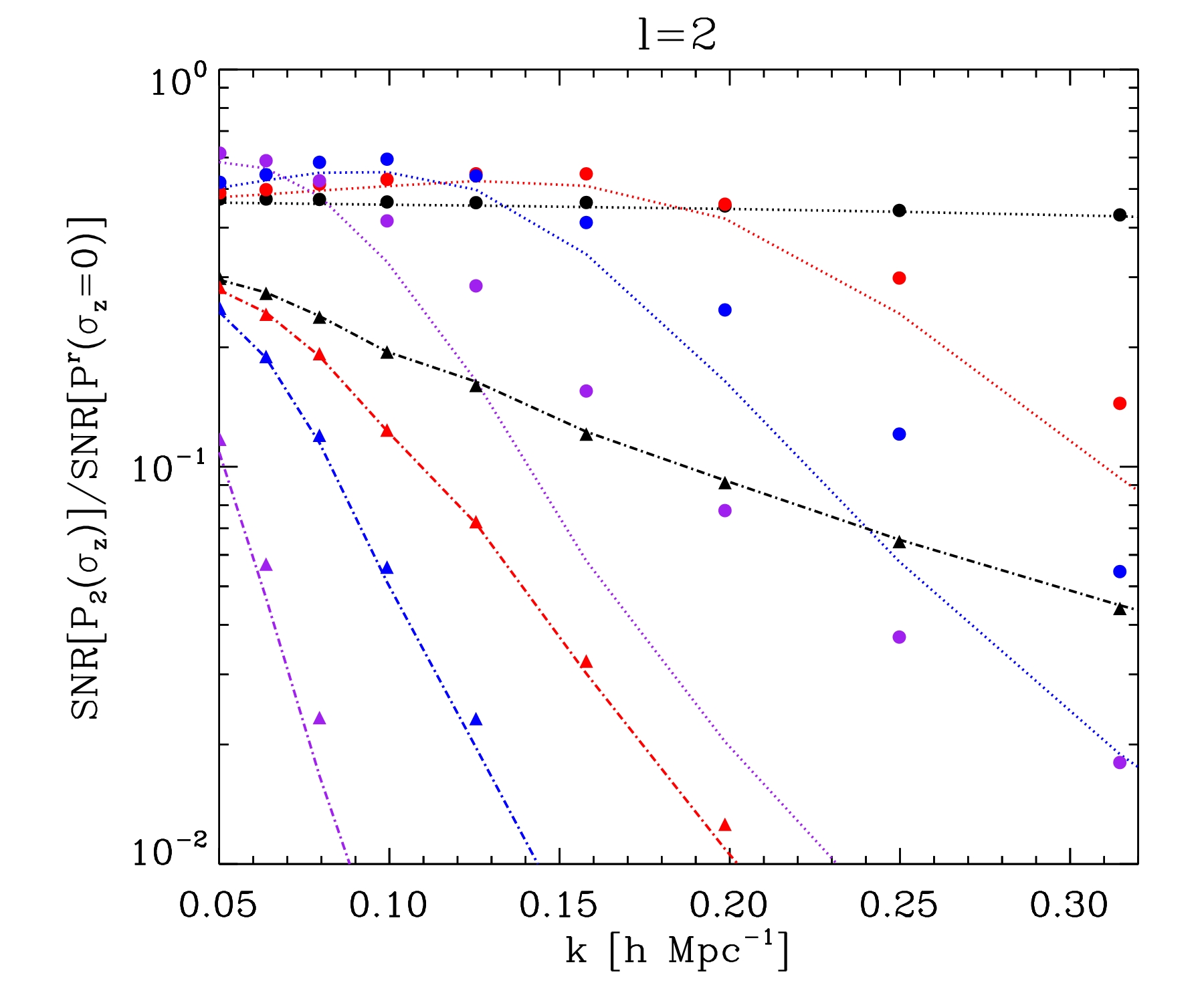}

\includegraphics[width=0.37\textwidth]{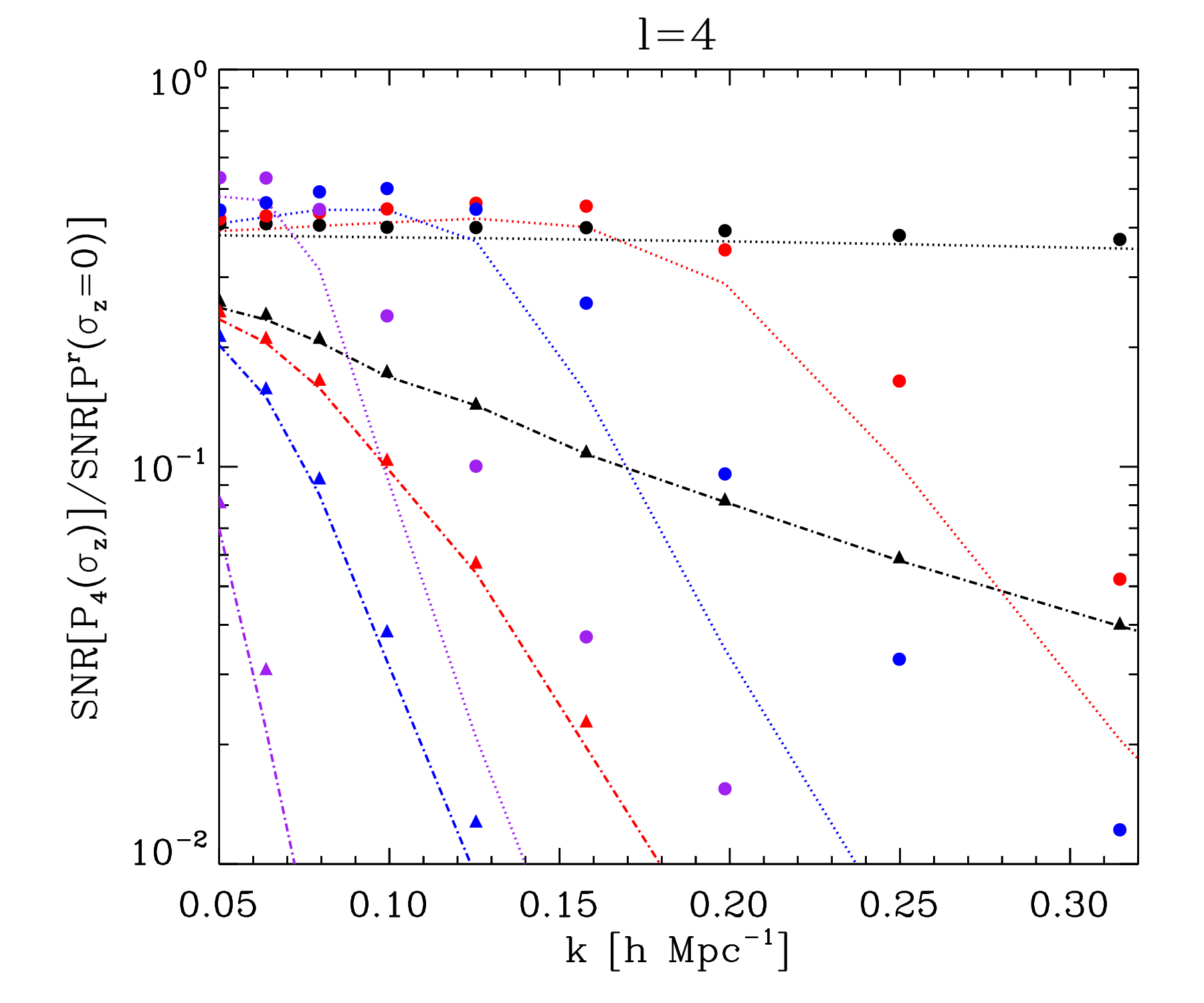}
\end{center}
\caption{
\label{fig:4}
Ratio of the ${\rm SNR}$ of power spectrum moments, $P_\ell/\sigma[P_\ell]$, to
that of a case with no photo-$z$ errors and $n=0.03\,h^3\,\Mpc^{-3}$ in real
space, $P^r/\sigma[P^r]$. On large scales, the ${\rm SNR}$ of $P_\ell$ is
greater for samples with large number densities and photo-$z$ errors. However,
on small scales the ${\rm SNR}$ of power spectrum moments decreases by increasing
$\sigma_z$.}
\end{figure}

From Eq.~\ref{eq:s2n_toym} we find that the scale corresponding to the local
maximum of the ${\rm SNR}$, $\partial ({\rm SNR}) / \partial (k\,\sigma_{\rm
eff})^2 = 0$, is $(k\,\sigma_{\rm eff})^2 = \text{ln}(\eta_{21})/\Delta \mu^2$.
That is $(k\,\sigma_{\rm eff})^2 \simeq 1$, as shown in the top panel of
Fig.~\ref{fig:4}.

\begin{table}
\begin{center}
\caption{Ratio of the total ${\rm SNR}$ of $P_0$, $P_2$, and $P_4$ and the
${\rm SNR}$ of the real-space power spectrum with no photo-$z$ errors and
$n=0.03\,h^3\,\Mpc^{-3}$. This ratio is computed from $0.05$ to
$0.3\,h\Mpc^{-1}$, the range of scales used for BAO analyses in \S\ref{sec:5o}.}
\label{tab:0}
\begin{tabular}{ccc}
$n\,[h^3\,\Mpc^{-3}]$&$\sigma_z\,[\%]$&
$r_{\rm SNR}=\frac{{\rm SNR}[\sigma_z]}{{\rm SNR}_r[\sigma_z=0]}$\\\hline\hline
$10^{-2}$&  0 & 1.00\\
$10^{-2}$&0.3 & 0.61\\
$10^{-2}$&0.5 & 0.52\\
$10^{-2}$&1.0 & 0.39\\\hline
$10^{-3}$&  0 & 0.70\\
$10^{-3}$&0.3 & 0.40\\
$10^{-3}$&0.5 & 0.31\\
$10^{-3}$&1.0 & 0.22\\\hline
$10^{-4}$&  0 & 0.21\\
$10^{-4}$&0.3 & 0.11\\
$10^{-4}$&0.5 & 0.08\\
$10^{-4}$&1.0 & 0.05\\\hline
\end{tabular}
\end{center}
\end{table}

\subsubsection{Comparison with simulations}

We now compare our analytic expressions for the ${\rm SNR}$ of power spectrum
moments (i.e. those derived in the previous two subsections) with the results
from the COLA ensemble. In Fig.~\ref{fig:4} we show the ${\rm SNR}$ of $P_0$,
$P_2$, and $P_4$ relative to that of the real-space power spectrum with
$n=0.03\,h^3\,\Mpc^{-3}$ and no photo-$z$ errors. We present the redshift-space
results for two number densities, as indicated by the legend. In all cases we
can see that our model, indicated by lines, approximately reproduces the
numerical data, displayed by symbols. The differences for
$n=10^{-2}h^3\,\Mpc^{-3}$ on large scales are driven by our model for the shot
noise subtraction. Independently of the size of photo-$z$ errors, on large
scales the ${\rm SNR}$ of each moment is lower than that of the real-space power
spectrum. This implies that, in the regime where shot noise is subdominant and
despite the clustering enhancement due to large-scale RSD, in redshift space the
${\rm SNR}$ of power spectrum moments is lower than that in real space. This
confirms the predictions of the toy model introduced in the previous section.

For samples with photo-$z$ errors, we appreciate an increase in the ${\rm SNR}$
relative to the case with no photo-$z$ errors on scales where $k\,\sigma_{\rm
eff} \simeq 1$, and a decrease on scales where the contribution of shot noise is
the dominant in Eq.~\ref{eq:smodel}. Moreover, for $\sigma_z \lesssim 0.5\,\%$,
the enhancement occurs on the scales where BAO are located. As BAO are
suppressed by the non-linear evolution of the matter density field and RSD, this
enhancement could imply that stronger cosmological constraints are derived from
samples with sub-percent photo-$z$ errors. We will return to this in the next
section. Note that these apparent advantages may disappear after applying
reconstruction procedures to the density field.

In Table \ref{tab:0} we show the ratio between the total ${\rm SNR}$ of $P_0$,
$P_2$, and $P_4$ (computed taking into account the covariances among them, see
Fig.~\ref{fig:3}) and that of the real-space power spectrum with no photo-$z$
errors and $n=0.03\,h^3\,\Mpc^{-3}$. This ratio, $r_{\rm SNR}$, is computed from
$0.05$ to $0.3\,h\Mpc^{-1}$, the range of scales that we use for BAO analyses in
\S\ref{sec:5o}. We find that $r_{\rm SNR}=1$ for samples with no photo-$z$
errors and $n=10^{-2}h^3\,\Mpc^{-3}$, and thus there is the same amount of
information in redshift space as in real space. Nevertheless, this information
is no equally distributed in both spaces. In real space and for samples with no
photo-$z$ errors all the information is stored in the $\ell=0$ moment, whereas
in redshift space it is distributed between the $\ell=0$, 2, and 4 moments. In
this case, we find that the $\ell=0$ moment accounts for $\simeq 97\,\%$ of the
signal, the $\ell=2$ moment the remaining $\simeq 3\,\%$, and the $\ell=4$
moment does not contain any new information. Examining samples with a lower
number density we find that the $\ell=0$ moment encodes even more information,
whereas $P_4$ starts to provide as much information as $P_2$. This motivates the
analysis of $P_0$, $P_2$, and $P_4$ in \S\ref{sec:5o}.

The results of Table \ref{tab:0} indicate that the value of $r_{\rm SNR}$
decreases by increasing the size of photo-$z$ errors. This is somewhat expected,
as we can see in Fig.~\ref{fig:4} that the increment in the ${\rm SNR}$ for
samples with photo-$z$ errors only occurs on large scales, whereas on small
scales the ${\rm SNR}$ is reduced.

\begin{figure}
\begin{center}
\includegraphics[width=0.475\textwidth]{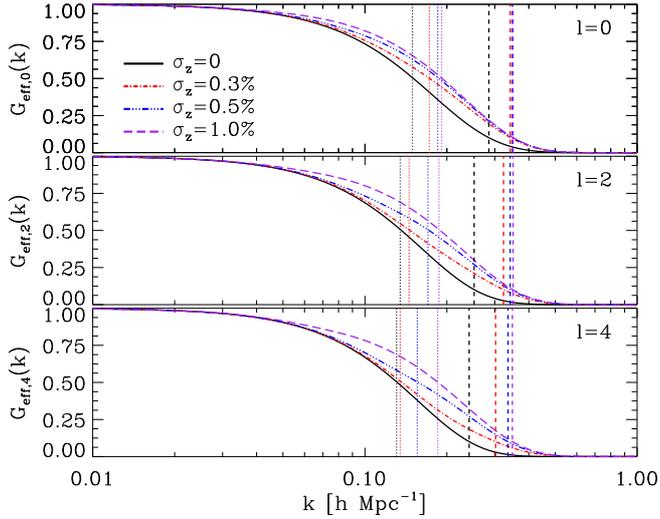}
\end{center}
\caption{
\label{fig:5}
Suppression of BAO due to the combined effect of photo-$z$ errors,
non-linearities, and RSD. The top, middle, and bottom panels show the results
for $P_0$, $P_2$, and $P_4$, respectively. Dotted and dashed lines indicate when
the suppression is greater than a factor of 2 and 10, respectively. The
suppression of BAO is weaker for samples with photo-$z$ errors because they
decrease the weight of $k$-modes parallel to the LOS in the angular average,
where in redshift space these modes are more suppressed than the perpendicular
ones owing to RSD.}
\end{figure}

\begin{figure}
\begin{center}
\includegraphics[width=0.375\textwidth]{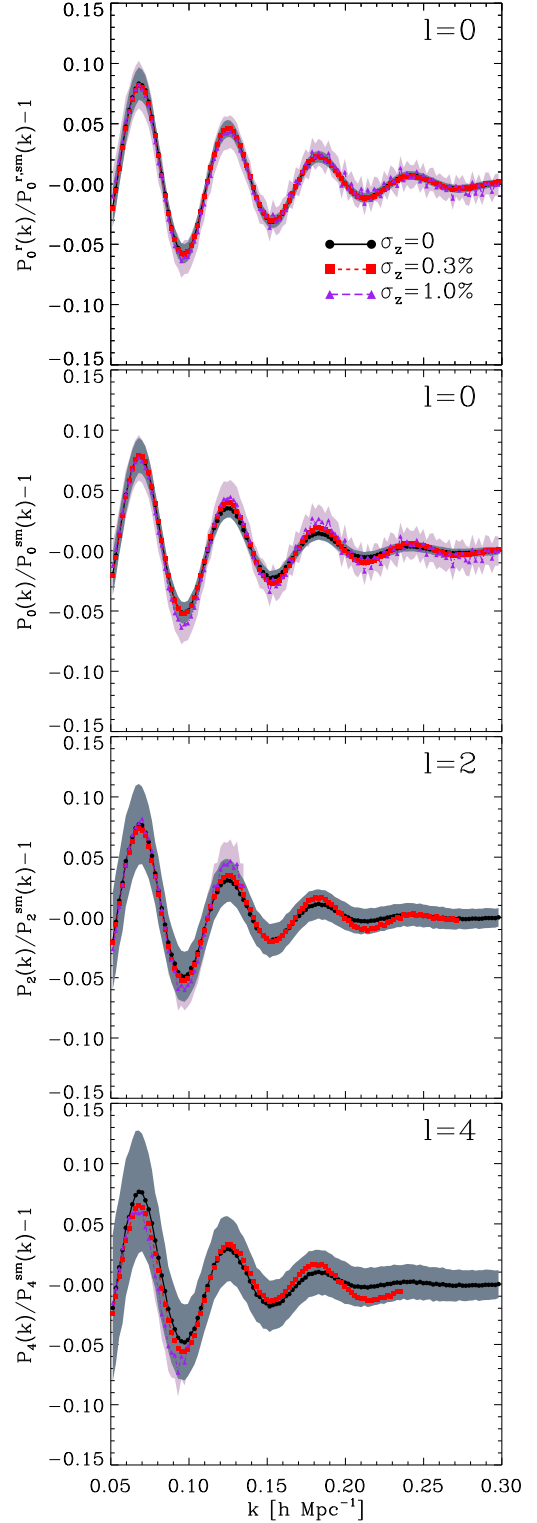}
\end{center}
\caption{
\label{fig:6}
Average $B_\ell$ computed using $1\,000$ samples from the COLA ensemble with
$n=0.01\,h^3\,\Mpc^{-3}$ for $P^r_0$, $P_0$, $P_2$, and $P_4$. The grey and
purple areas denote the $1\sigma$ confidence region for samples with $\sigma_z =
0$ and $1\,\%$, respectively. For samples with photo-$z$ errors we do not show
the results up to $k \sim 0.3\,h\,\Mpc^{-1}$ because our procedure for
extracting $B_\ell$ does not work well on scales where power spectrum moments
are dominated by shot noise. In real space the amplitude of BAO is not modified
by photo-$z$ errors, whereas in redshift space it grows with their size. This
confirms the predictions of Fig.\ref{fig:5}.}
\end{figure}

\section{Effect of photometric redshift errors on BAO} 
\label{sec:3o}

In this section we investigate the effect of photo-$z$ errors on the BAO 
feature imprinted in power spectrum moments. In \S\ref{sec:3a} we study the
suppression of BAO due to the combined effect of photo-$z$ errors, the 
non-linear evolution of the matter density field, and RSD. In \S\ref{sec:3b}
we analyse the cosmological information encoded in BAO, in \S\ref{sec:3c} we
introduce a model to estimate the uncertainty in measuring the BAO scale, and
in \S\ref{sec:3d} we address how to extract cosmological information from the
joint analysis of power spectrum moments.

\subsection{The shape of the BAO signal}\label{sec:3a}

Let us begin by considering the following quantity:

\begin{equation}
B_\ell(k) \equiv \frac{P_\ell(k)}{P_\ell^{\rm sm}(k)} - 1,
\end{equation}

\noindent where $P_\ell^{\rm sm}$ is a no-wiggle version of $P_\ell$. This is 
a quantity with the same broadband shape as $P_\ell$ but no BAO. Therefore, 
$B_\ell$ is insensitive to the overall shape of the observed moments and isolates
the BAO wiggles.

Motivated by Renormalized Perturbation Theory \citep{Crocce2008}, we write
the non-linear redshift-space power spectrum as:

\begin{equation}
\label{eq:pkmug}
P(k,\mu) = \left[P_{\rm lin}(k)\,G(k,\mu) + P_{\rm mc}(k,\mu) \right]\,b^2\,\f^2,
\end{equation}

\noindent where $P_{\rm lin}$ is the linear theory power spectrum in real
space, $P_{\rm mc}$ denotes contribution of the coupling between different 
$k$-modes, and $G$ is a propagator that controls the suppression of BAO 
due to non-linearities and RSD. This propagator is well approximated by a 
2D exponential function:

\begin{equation}
G(k,\mu) = \exp\bigg\{-\left[ (1 - \mu^2)\,k^2\,\sigma^2_{\perp} + 
\mu^2\,k^2 \sigma^2_{\parallel}\right] \bigg\},
\end{equation}

\noindent where $\sigma_{\parallel}$ and $\sigma_{\perp}$ are parameters that
control the suppression of $k$-modes along and perpendicular to the LOS,
respectively. Note that, in redshift space $\sigma_{\parallel}>\sigma_{\perp}$.
This is a consequence of peculiar velocities and the non-linear mapping between
real and redshift-space positions, and it implies that the larger the value of
$\mu$, the greater the suppression of BAO for a given wavemode
\citep[e.g.,][]{SeoEisenstein2007, Sanchez2008}. 

Let us now write a theoretical model for $B_\ell$:

\begin{equation}
B_\ell(k) \simeq B_{\rm lin}(k)\,G_{\rm eff,\ell}(k),\\
\end{equation}

\noindent where $B_{\rm lin}=P_{\rm lin}/P^{\rm sm}_{\rm lin}-1$, and

\begin{equation}
G_{\rm eff,\ell}(k) = \frac{\langle G(k,\mu)\,\mu^\ell
\,\f^2\rangle_{\hat{\mathbf{k}}}} {\langle\mu^\ell
\,\f^2\rangle_{\hat{\mathbf{k}}}}, \label{eq:geff2}
\end{equation}

\noindent In deriving these expressions, we have assumed that $P\simeq[P_{\rm
lin}^{\rm sm} + (P_{\rm lin}-P_{\rm lin}^{\rm sm})G]b^2\f^2$ and $P^{\rm
sm}\simeq P_{\rm lin}^{\rm sm} b^2\f^2$. These assumptions are not strictly
correct because they do not take into account the $P_{\rm mc}$ term in
Eq.~\ref{eq:pkmug}, and thus Eq.~\ref{eq:geff2} does not predict a shift in the
BAO scale due to non-linearities as the full model does \citep{Crocce2008}. 

The term $G_{\rm eff,\ell}$ gives the effective suppression of BAO in $P_\ell$,
as a function of the scale. As we can see, this quantity is a weighted average
of $G(k,\mu)$, where the weights are given by photo-$z$ errors and RSD.
Therefore, a balance between these two aspects will determine the final
appearance of BAO, as we will see next.

\subsubsection{The Gaussian case and comparison with simulations}
\label{sec:3ab}

In the particular case of Gaussian photo-$z$ errors, $G_{\rm eff,\ell}$ has an
analytic expression. In the top, middle, and bottom panels of
Fig.~\ref{fig:5} we show $G_{\rm eff,0}$, $G_{\rm eff,2}$, and $G_{\rm
eff,4}$, respectively. To build this figure we used
$\sigma_{\parallel}=6.8\,h^{-1}\,\Mpc$ and $\sigma_{\perp}=4.3\,h^{-1}\,\Mpc$
(these values were set by our fits to the numerical simulations presented in
\S\ref{sec:5o}). The vertical dotted and dashed lines indicate the scale at
which the suppression of BAO is $50\,\%$ and $90\,\%$, respectively.

As we can see, the BAO suppression is {\it weaker} for samples with larger
photo-$z$ errors. This has an interesting consequence, {\it photo-$z$ errors
make BAO wiggles to appear sharper}. Furthermore, for a given photo-$z$ error,
the suppression is stronger for higher order moments. This counter-intuitive
result can be understood by recalling that LOS modes -- where BAO smearing is
more significant -- contribute less to a given moment, thus, when photo-$z$ are
included, BAO appear more alike to the less-damped real-space case.

In order to check the accuracy of the predictions of Eq.~\ref{eq:geff2}, we
measure $B$ from the COLA ensemble. To obtain the no-wiggle power spectrum
$P_\ell^{\rm sm}$, we fit $P_\ell$ using the following model

\begin{align}
P_{\rm fit, \ell}(k) =& A_{0, \ell} \langle \mu^\ell \,\f^2\rangle_{\hat{\mathbf{k}}} 
P_{\rm nw}(k) + A_{1, \ell} k^2 + A_{2, \ell} k + A_{3, \ell} 
\nonumber\\
&+ A_{4, \ell} k^{-1} + A_{5, \ell} k^{-2} + A_{6, \ell} k^{-3},
\end{align}

\noindent where $P_{\rm nw}$ is the non-wiggle power spectrum from
\citep{Eisenstein1998}, and the factors $A_{i,\ell}$ give us enough freedom to
fit the broadband shape of $P_\ell$ without fitting the oscillations. We
explicitly show this in Fig.~\ref{fig:6}, where we display the average value of
$B$ measured from samples of the COLA ensemble with $n=10^{-2}h^3\,\Mpc^{-3}$.
In descending order, the panels show the results for $P^r_0$, $P_0$, $P_2$, and
$P_4$. Note that for samples with sub-percent photo-$z$ errors, $B_2$ and $B_4$
are not displayed up to $k \sim 0.3\,h\,\Mpc^{-1}$. This is because on small
scales the measurements start to be progressively dominated by shot noise.

We find that in real space the BAO feature is the identical in the cases with
and without photo-$z$ errors. This is expected from Eq.~\ref{eq:geff2}, as in
real space $\sigma_{\parallel} = \sigma_{\perp}$, and thus $G_{\rm eff,0} =
e^{-(k\,\sigma_\perp)^2}$. However, in redshift space, BAO are less suppressed
for samples with greater photo-$z$ errors, confirming the predictions of
Fig.~\ref{fig:5} and our analytic model.


\subsection{Cosmological information encoded in BAO} 
\label{sec:3b}

We now explore the cosmological information encoded in the BAO feature. Let us
consider a given scale $k = \sqrt{k_{\parallel}^2 + k_{\perp}^2}$ and angle 
$\mu=k_{\parallel}/k$ in the two-dimensional power spectrum. Assuming a
fiducial cosmology, they are observed as

\begin{equation}
k^{\rm fid} = k/\alpha \equiv k\sqrt{\mu^2 \alpha_\parallel^{-2} +
(1-\mu^2) \alpha_\perp^{-2}},
\end{equation}

\begin{equation}
\mu^{\rm fid} = \epsilon \mu \equiv 
\frac{\mu}{\sqrt{\mu^2+(1-\mu^2)(\alpha_\parallel/\alpha_\perp)^2}},
\end{equation}

\noindent where $\alpha_{\parallel}\equiv \frac{H^{\rm fid}(z)r_s^{\rm
fid}}{H(z)r_s}$ and $\alpha_{\perp}\equiv \frac{D_{A}(z)r_s^{\rm fid}}{D^{\rm
fid}_{A}(z)r_s}$. In the above expressions, $r_s$ is the sound horizon scale,
$D_A$ is the angular diameter distance, $H$ is the Hubble parameter, and ${\rm
fid}$ denotes these quantities in the fiducial cosmology. The observed
redshift-space power spectrum moments are thus

\begin{align}
P_\ell(k/\alpha_{\ell}) =& \alpha_\parallel^{-1} \alpha_\perp^{-2}
\bigg\langle \epsilon^{\ell+1} \mu^\ell 
\left[1+\epsilon^2\mu^2\left(\alpha_\parallel^2 \alpha_\perp^{-2}-1\right)\right]
\nonumber\\
&\mathcal{F}^2(k/\alpha, \epsilon \mu) P_0^r(k/\alpha) 
\bigg\rangle_{\hat{\vk}},
\end{align}

\noindent where assuming an incorrect cosmology would cause isotropic and
anisotropic deformations. For the full expression of deformations in observed
multipoles see \citet{Padmanabhan08} in Fourier space and \citet{Xu13} in
configuration space. 

Since here we are mostly interested in the information encoded in the BAO scale,
we follow \citet[][R15 hereafter]{Ross2015} and focus only on the stretch
parameters $\alpha_\ell$ as a function of $\alpha_\parallel$ and
$\alpha_\perp$\footnote{In what follows we will assume that
$\alpha_\parallel^2/\alpha_\perp^2=1$ and thus $\epsilon=1$.}. As in R15, we
will also assume that the information on $\alpha_\ell$ is separable from the
overall shape of power spectrum moments and that the information in different
$\mu$ bins is independent. Under these assumptions:

\begin{equation}
\alpha_\ell(k) = \frac{\left\langle \,\mu^\ell\,\f^2 \alpha \right
\rangle_{\hat{\vk}}}{\langle \mu^\ell \f^2 \rangle_{\hat{\vk}}},
\end{equation}

\noindent where the scale-dependence of $\alpha_\ell$ emerges from the
scale-dependence of $\f^2$. The previous expression reduces to Eq.~6 of R15 for
samples with no photo-$z$ errors nor small-scale RSD. Therefore, small-scale RSD
and/or photo-$z$ errors induce a scale-dependence in the stretch parameter.

As in R15, we find that the first order expansion of $\alpha_\ell$ around the
fiducial solution can be expressed as $\alpha_\ell(k) =
\alpha^{m_\ell(k)}_{\parallel} \alpha^{n_\ell(k)}_{\perp}$, where $m_\ell$ and
$n_\ell$ are given by:

\begin{align}
\label{eq:m}
m_\ell(k) \equiv& \left.\frac{\partial\left<\alpha_\ell\right>}
{\partial\alpha_{\parallel}}\right|_{\alpha_\parallel=\alpha_\perp=1} 
= \frac{\langle \mu^{\ell+2} \f^2 \rangle_{\hat{\vk}}}
{\langle \mu^\ell \f^2 \rangle_{\hat{\vk}}},\\
\label{eq:n}
n_\ell(k) \equiv& \left.\frac{\partial\left<\alpha_\ell\right>}
{\partial\alpha_{\perp}}\right|_{\alpha_\parallel=\alpha_\perp=1} 
=1- m_\ell(k),
\end{align}

\noindent where the higher the value of $m_\ell$, the more sensitive
$\alpha_\ell$ is to the Hubble parameter. For the case of Gaussian photo-$z$
errors, $m_\ell$ and $n_\ell$ have analytic expressions.

Going back to the stretch parameter, the known case with $m_0=1/3$ and $n_0=2/3$
\citep{Eisenstein2005} is only recovered in real space without photo-$z$ errors.
In redshift space, there is a dependence of $m_\ell$ and $n_\ell$ on $\beta$
even if $\sigma_z = 0$. In general, the effect of photo-$z$ errors and
small-scale RSD is to decrease the sensitivity of $\alpha_\ell$ on $H$, whereas
large-scale RSD have the opposite effect. Nonetheless, the exact degeneracy
between $\alpha_{\parallel}$ and $\alpha_{\perp}$ also depends on the properties
of analysed sample, such as its large-scale bias.


\subsection{Toy model for the uncertainty in the stretch parameter} 
\label{sec:3c}

In \S\ref{sec:2c} we studied the ${\rm SNR}$ of power spectrum moments and in
the previous section we showed that the suppression of the BAO feature depends
on $G_{\rm eff,\ell}$. In this section we use this to derive an analytic
estimation of the uncertainty in $\alpha_\ell$ as a function of large-scale
bias, number density, photo-$z$ error, and cosmology.

To build an estimator for the precision measuring $\alpha_\ell$, we will assume
that its uncertainty is given by the convolution of the one measuring $P_\ell$
and the amplitude of BAO wiggles in $P_\ell$. The latter obviously depends on
the suppression of BAO due to non-linearities and RSD, which as we explained
before is captured by $G_{\rm eff,\ell}$. For conventional approaches using
Fisher matrix forecast see for instance \citet{SeoEisenstein2003}.

We estimate the amplitude of unsuppressed BAO in linear theory by measuring the
absolute value of the local extrema of $1+B_{\rm lin}$. Then, we do a linear fit
of these values, $B_{\rm amp}$. The product of this quantity and
Eq.~\ref{eq:geff2}, $G_{{\rm eff},\ell}\,B_{\rm amp}$, gives thus the maximum
amplitude of BAO wiggles in $P_\ell$ under the presence of non-linearities, RSD,
and photo-$z$ errors.

In BAO analyses, the stretch parameter $\alpha_\ell$ is extracted from an
interval of scales. As this parameter is scale-dependent due to small-scale
velocities and photo-$z$ errors, it is useful to define an effective stretch
parameter $\alpha_{{\rm eff},\ell}$. We estimate the uncertainty in this
parameter as the uncertainty in a moment times the amplitude of BAO wiggles

\begin{equation}
\label{eq:intsneff}
\hat{\sigma}[\alpha_{{\rm eff},\ell}] = \frac{A_\ell}{(k_{\rm max}-k_{\rm min})^c} 
\left(\int_{k_{\rm min}}^{k_{\rm max}}
\text{d}k\,\frac{P_\ell^2\,G_{\rm eff,\ell}^2\,B_{\rm
amp}^2}{\sigma^2[P_\ell]}\right)^{-0.5},
\end{equation}

\noindent where $A_\ell$ and $c$ are free parameters. Note that we assumed that
the off-diagonal terms of the covariance matrices of power spectrum moments are
negligible on the range of scales where BAO are located (see \S\ref{sec:2bd}).



\begin{figure}
\begin{center}
\includegraphics[width=0.475\textwidth]{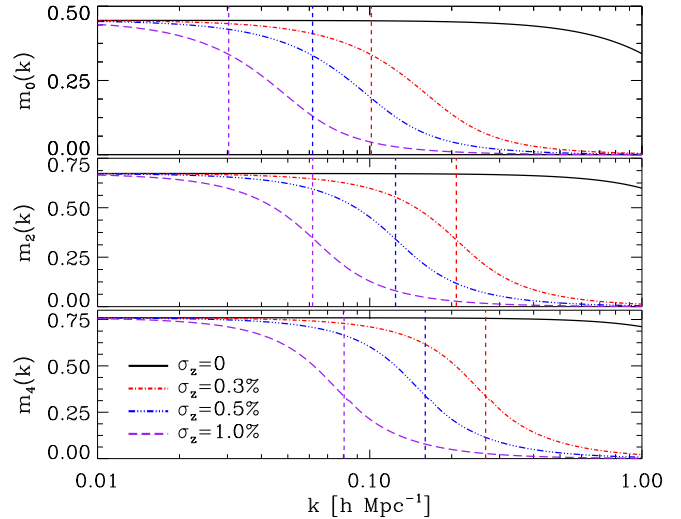}
\end{center}
\caption{
\label{fig:7} 
Sensitivity of BAO to the Hubble parameter as function of the scale, $m_\ell$.
The top, medium, and bottom panels show the results for BAO in $P_0$, $P_2$, and
$P_4$, respectively. The vertical lines indicate the scale at which the
dependence of BAO on the Hubble parameter is smaller than for samples with no
photo-$z$ errors in real space ($m = 1/3$). As expected, the larger the
photo-$z$ error, the smaller is the dependence of BAO on the Hubble parameter. 
We can see that samples with no photo-$z$ errors also present a dependence of
$m_\ell$ on the scale, which occurs due to small-scale velocities.
}
\end{figure}


\subsection{The scale-dependence of cosmological information}
\label{sec:3d}

In \S\ref{sec:3b} we showed that there is a scale dependence for the cosmological
information encoded in BAO, which introduces an additional complication while
extracting information from BAO analyses. We present in the top, middle, and
bottom panel of Fig.~\ref{fig:7} the value of $m_0$, $m_2$, and $m_4$ as a
function of the scale, respectively. Solid lines indicate the results for
different photo-$z$ errors, $b=1$, and our adopted cosmology (c.f. \S2.1). Dashed
lines denote when the sensitivity of BAO to the Hubble parameter is the same as in
real-space, $m=1/3$. As we can see, the higher is the order of the moment, the
greater is the sensitivity on the Hubble parameter. For example, if we look at
samples with $\sigma_z=0.3\,\%$, $m_0$, $m_2$, and $m_4$ are greater than $1/3$
up to $k\simeq 0.1$, $0.2$, and $0.3\,h\,\Mpc^{-1}$, respectively. Therefore,
the analysis of BAO in the $\ell=4$ moment of samples with $\sigma_z=0.3\,\%$ is
more sensitive to the Hubble parameter than in real space. This highlights the
importance of not only analysing BAO in the $l=0$ and $l=2$ moments. In
addition, we can see that there is a scale dependence of $m_\ell$ for samples
with no photo-$z$ errors, which is caused by small-scale velocities.

We can derive the relation between the precision in measuring $\alpha_\ell$
and the radial and perpendicular components of $\alpha$ -- Hubble parameter 
and angular diameter distance, respectively -- using Eqs.~\ref{eq:m} and 
\ref{eq:n}:

\begin{equation}
\sigma^2[\alpha_{\ell,\ell'}]=m_\ell^2\,\sigma^2[\alpha_{\parallel}]
+n_{\ell'}^2\,\sigma^2[\alpha_{\perp}],
\end{equation}

\noindent where the uncertainty in the radial and perpendicular components are
$\sigma[\alpha_{\parallel}]$ and $\sigma[\alpha_{\perp}]$, respectively.
Consequently, to extract the precision in measuring both components we need to
perform a joint fit of at least two power spectrum moments.

In the same way that we computed an effective value for the stretch parameter
in the previous section, we can now estimate an overall degeneracy between 
the parallel and perpendicular components of $\alpha_\ell$ when they are
estimated from an interval of scales, $m_{\rm eff,\ell}$. To do this, we
compute the variance-weighted average of $m_\ell$ and $n_\ell$ over the
desired $k$-interval, where the variances are given by Eq.~\ref{eq:intsneff}. 
Explicitly,

\begin{align}
\label{eq:meff}
m_{\rm eff,\ell} =& \frac{\int_{k_{\rm min}}^{k_{\rm max}} 
\text{d}k\, \frac{m_\ell P_\ell^2 G_{\rm eff,\ell}^2 B_{\rm amp}^2}
{\sigma^2[P_\ell]}}
{\int_{k_{\rm min}}^{k_{\rm max}} \text{d}k\,
\frac{P_\ell^2 G_{\rm eff,\ell}^2 B_{\rm amp}^2}{\sigma^2[P_\ell]}},\\
\label{eq:neff}
n_{\rm eff,\ell} =& 1-m_{\rm eff,\ell},
\end{align}

\noindent and thus, the degeneracy between the overall precision in the parallel
and perpendicular components of $\alpha_{\rm eff,\ell}$ is given by

\begin{equation}
\label{eq:degmn}
\sigma^2[\alpha_{\rm eff,\ell}]=m_{\rm eff,\ell}^2
\,\sigma^2[\alpha_{\parallel}]
+n_{\rm eff,\ell}^2\,\sigma^2[\alpha_{\perp}].
\end{equation}

Nevertheless, we are interested in the precision measuring $H$ and $D_A$. In
\S\ref{sec:5o} we conduct a joint analysis of BAO in $P_0$, $P_2$, and $P_4$ to
compute their uncertainties, as it is required the analysis of at least two
moments to break the degeneracy in cosmological information along and
perpendicular to the LOS. From now on, we drop the subindex ${\rm eff}$ for
simplicity. In this way,

\begin{align}
\mathbf{\Sigma} = & \mathbf{M}^+\,\mathbf{E}\,(\mathbf{M}^+)^{\rm T},\\
\mathbf{\Sigma} = &
\begin{pmatrix}
  \sigma^2[\alpha_\parallel] & \sigma[\alpha_{\parallel,\perp}]\\
  \sigma[\alpha_{\parallel,\perp}] & \sigma^2[\alpha_\perp]\\
\end{pmatrix},\\
\mathbf{M} = &
\begin{pmatrix}
  m_0 & n_0 \\
  m_2 & n_2 \\
  m_4 & n_4 \\
 \end{pmatrix},\\
\mathbf{E} = &
\begin{pmatrix}
  \sigma^2[\alpha_0] & \sigma[\alpha_{0,2}] & \sigma[\alpha_{0,4}] \\
  \sigma[\alpha_{0,2}] & \sigma^2[\alpha_2] & \sigma[\alpha_{2,4}] \\
  \sigma[\alpha_{0,4}] & \sigma[\alpha_{2,4}]  & \sigma^2[\alpha_4] \\
 \end{pmatrix},
\end{align}

\noindent where $\mathbf{M}^+$ is the pseudoinverse of $\mathbf{M}$
\citep{Moore1920, Bjerhammar51, Penrose55}, $\mathbf{\Sigma}$ is the covariance
matrix of $H$ and $D_A$, and $\mathbf{E}$ is computed from BAO analysis of power
spectrum moments.



\section{Extracting information from BAO} \label{sec:4o}

In the previous sections we showed how photo-$z$ errors modify the amplitude
of power spectrum moments, their variances, the suppression of BAO, and the
cosmological information encoded in them. In this section we employ all this
information to create a model to unbiasedly extract the BAO scale, $\alpha$,
from observational and/or simulated data, even under the presence of photo-$z$
errors. We will employ this model on simulated catalogues in \S6.


\subsection{Modelling power spectrum moments}
\label{sec:4a}

Based on the expressions provided in \S\ref{sec:3a}, we can write the
following four parameter model for power spectrum moments:

\begin{equation}
\label{eq:model}
P_{T,\ell} = P^{\rm sm}_\ell(k) 
\left[B_{\rm lin}(k/\alpha_\ell) G_\ell(k/\alpha_\ell, 
\sigma_{\rm eff},\sigma_\perp,f) + 1\right]
\end{equation}

\noindent where $P^{\rm sm}_\ell$ is computed using the same approach as in
\S\ref{sec:3a}, the parameter $\alpha_\ell$ allows for stretching of the BAO,
$\sigma_\perp$ and $f=\sigma_\parallel/\sigma_\perp$ control BAO suppression.
Photo-$z$ errors enter only through $G_\ell$. Note that all these parameters 
only appear in the expression for $B_{\rm lin}$ and $G_\ell$, and thus they 
are only constrained by BAO information, i.e. our model extracts cosmological
information regardless of the overall shape of the analysed moment.

In the following sections we will fix $\sigma_{\rm eff}$ to the correct value.
We do this to break a degeneracy between $\sigma_\perp$ and $\sigma_{\rm
eff}$, as both control the suppression of BAO. In any case, we checked that
if include both parameters are let them vary, we recover the same results for
$\alpha_\ell$.


\subsection{Parameter Likelihood Calculation} \label{sec:4b}

To extract the information encoded in BAO, we jointly fit
$P_0$, $P_2$, and $P_4$. For this, we assume that the probability of
observing $\mathbf{d} = [P_0,P_2,P_4]$ is given by a multivariate Gaussian
distribution:

\begin{equation} 
\label{eq:likelihood}
\Pr(\mathbf{d}|\mathbf{\pi}) \propto \exp
\left[-\frac{1}{2}\left(\mathbf{d} - \mathbf{t}\right)^{\rm T}
\mathbf{C}^{-1}\left(\mathbf{d} - \mathbf{t}\right)\right],
\end{equation}

\noindent where $\mathbf{t}(k,\pi)=[P_{T,0},P_{T,2},P_{T,4}]$, and $\pi = \{
\alpha_0, \alpha_2, \alpha_4, \sigma_\perp, f\}$ denote the five free parameters
of our model. The priors on these parameters are assumed to be flat over the
range: $\alpha_\ell \in [0.75, 1.25]$ and $\sigma_\perp \in [0.5,10]$, and $f
\in [1, 10]$. We set the minimum value of $f$ to be 1 because in redshift space
the combination of RSD and non-linearities make $k$-modes along the LOS to be
the most suppressed (see \S\ref{sec:3a}). We note that the results do not change
if we set wider priors. $\mathbf{C}^{-1}$ is the data precision
matrix of $\mathbf{d}$, which we compute from the COLA ensemble as described in
\S\ref{sec:1b}. The interval of scales considered is $k =
[0.05-0.30]\,h\,\Mpc^{-1}$.

We sample the posterior probability distribution function of $\pi$ employing
the publicly available code {\tt emcee} \citep{Foreman-Mackey2013}. This code
is an affine invariant MCMC ensemble sampler that has been widely tested and
used in multiple scientific studies. We configure the code to analyse 
$\mathbf{d}$ using a chain of $100$ random walkers with $5\,000$ steps each,
and a burn-in phase of $500$ steps. We checked that this burn-in phase is
sufficient to obtain well-behaved chains.

Additionally, we checked that the standard deviations of the best-fit values
from the COLA ensemble are compatible with the uncertainties estimated from
the likelihood of each simulated catalogue.

\section{Results from simulated catalogues}
\label{sec:5o}

In this section we start by studying the quality of the BAO fitting model
introduced in the previous section. Then, we apply this model to simulated catalogues 
with different photo-$z$ errors and number densities. In addition, we study whether 
introducing photo-$z$ errors following PDFs different from Gaussian impacts our
results. Finally, we present a simple procedure to undo the effect of peculiar
velocities on power spectrum moments.

We note that in this section we analyse power spectrum moments computed from DM
particles. In principle, this could introduce a shift in the stretch parameter
with respect to computing power spectrum moments from haloes/galaxies. However,
\citet{Angulo2014} addressed this by measuring the stretch parameter from the
real- and redshift-space monopole of galaxies selected by star formation and
stellar mass in the Millennium XXL simulation \citep{Angulo2012}, finding no
additional shifts with respect the stretch parameter measured from DM particles.
Therefore, we will restrict our analysis to DM particles.


\begin{figure}
\begin{center}
\includegraphics[width=0.475\textwidth]{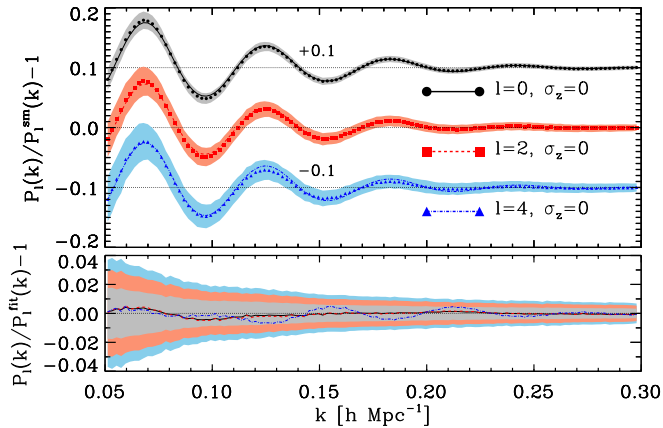}
\end{center}
\caption{
\label{fig:bkfit}
Relative difference between $P_0$, $P_2$, $P_4$ and their no-wiggle versions for
samples with no photo-$z$ errors and $n=10^{-2}\,h^3\,\Mpc^{-3}$. Symbols and
lines display the average results from the COLA ensemble and the average
best-fit to each mock using Eq.~\ref{eq:model}, respectively. Coloured regions
indicate the $1\sigma$ region of the scatter from mock-to-mock. The bottom panel
shows the relative difference between the average data and model. The precision
of our template is to within $1\,\%$, which is greater than the typical scatter
from mock-to-mock.}
\end{figure}


\subsection{Quality of our model} \label{sec:5a}

In Fig.~\ref{fig:bkfit} we show the result of fitting the model introduced in
Eq.~\ref{eq:model} to samples from the COLA ensemble with $\sigma_z=0$ and
$n=10^{-2}\,h^3\,\Mpc^{-3}$. Symbols indicate the average value of $B_\ell$
computed from the COLA ensemble, and lines the average of the best-fit to each
mock. We display the results for $\ell=0$, $\ell=2$, and $\ell=4$ in black, red,
and blue, respectively, where they are offset for clarity. Shaded areas enclose
the $1\sigma$ region from mock-to-mock. In the bottom panel we plot the relative
difference between the model and data, where we can see that in all cases the
typical deviations are statistically insignificant. In addition, our model
reproduces the simulated data to within $1\,\%$ on the scales shown, which is
more than enough for the next generation of galaxy surveys. We also checked that
we obtain similar results for samples with sub-percent photo-$z$ errors.

\begin{figure*}
\includegraphics[width=0.475\textwidth]{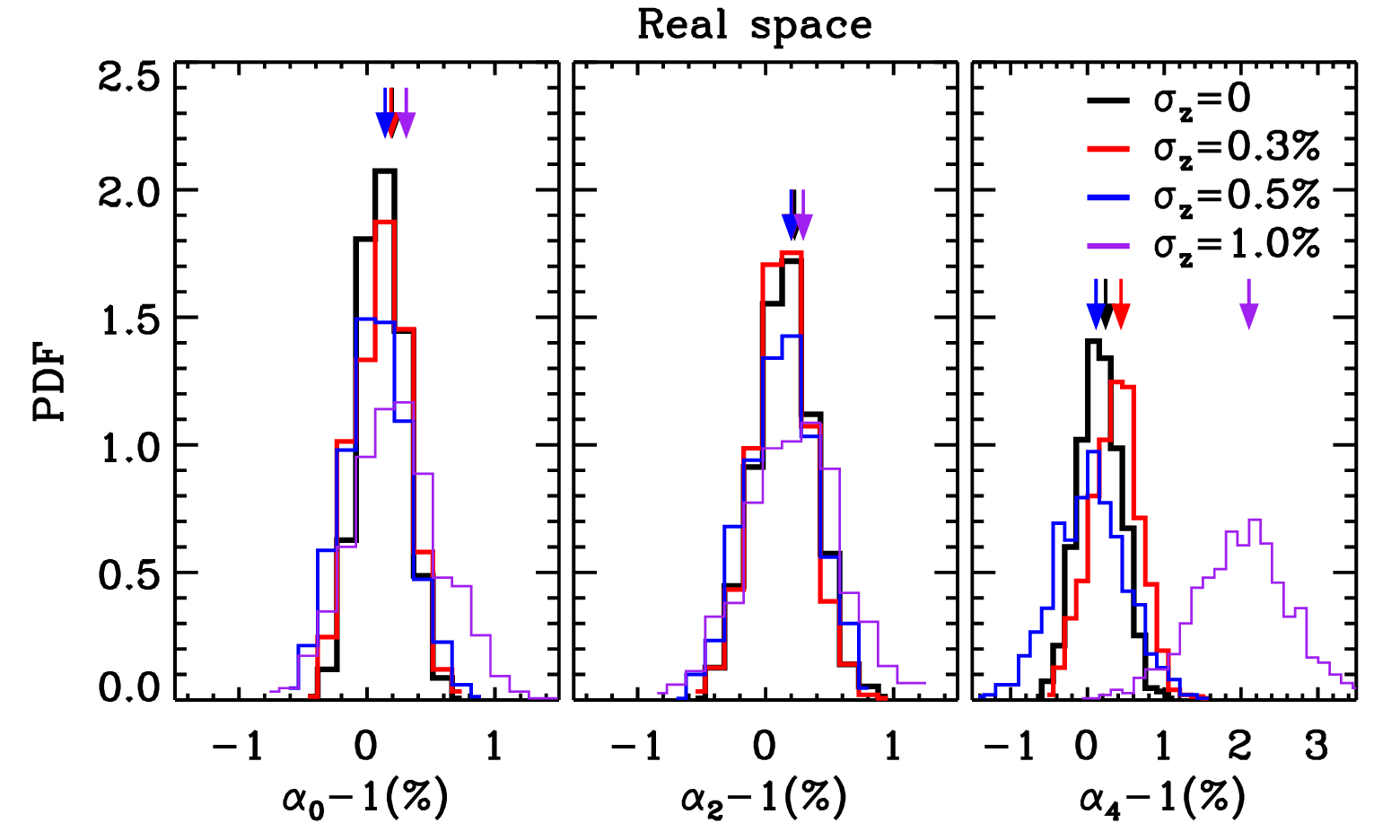}\includegraphics[width=0.475\textwidth]{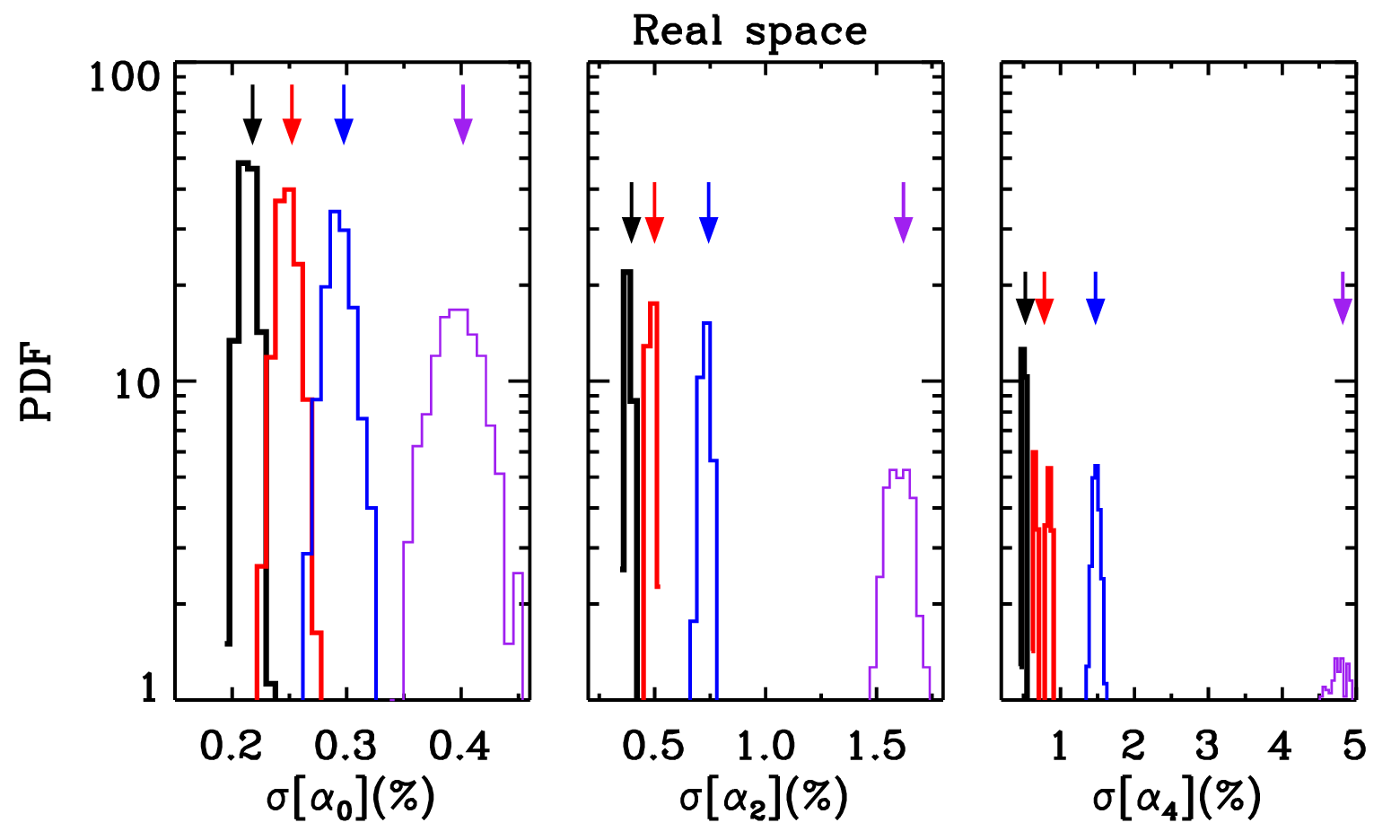}

\includegraphics[width=0.475\textwidth]{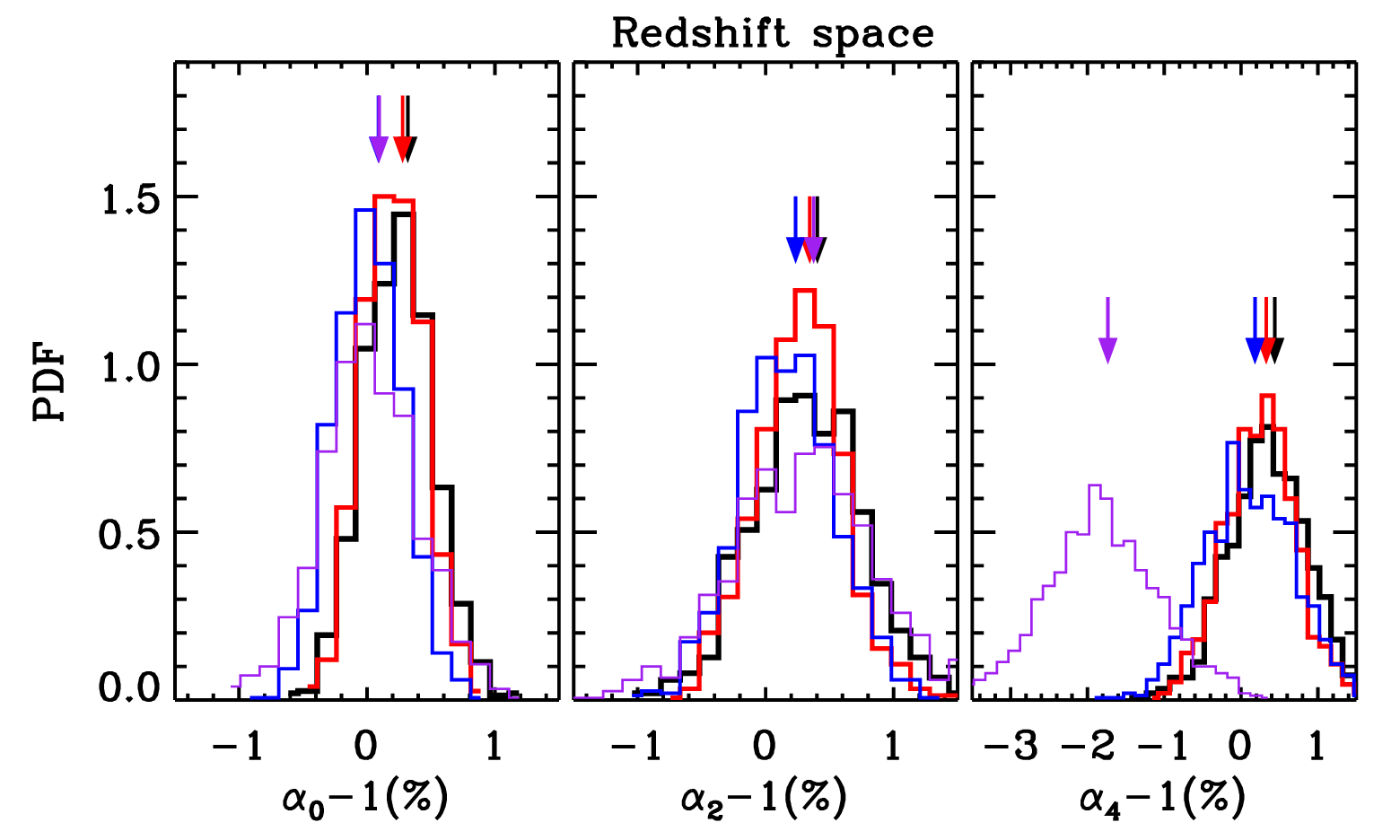}\includegraphics[width=0.475\textwidth]{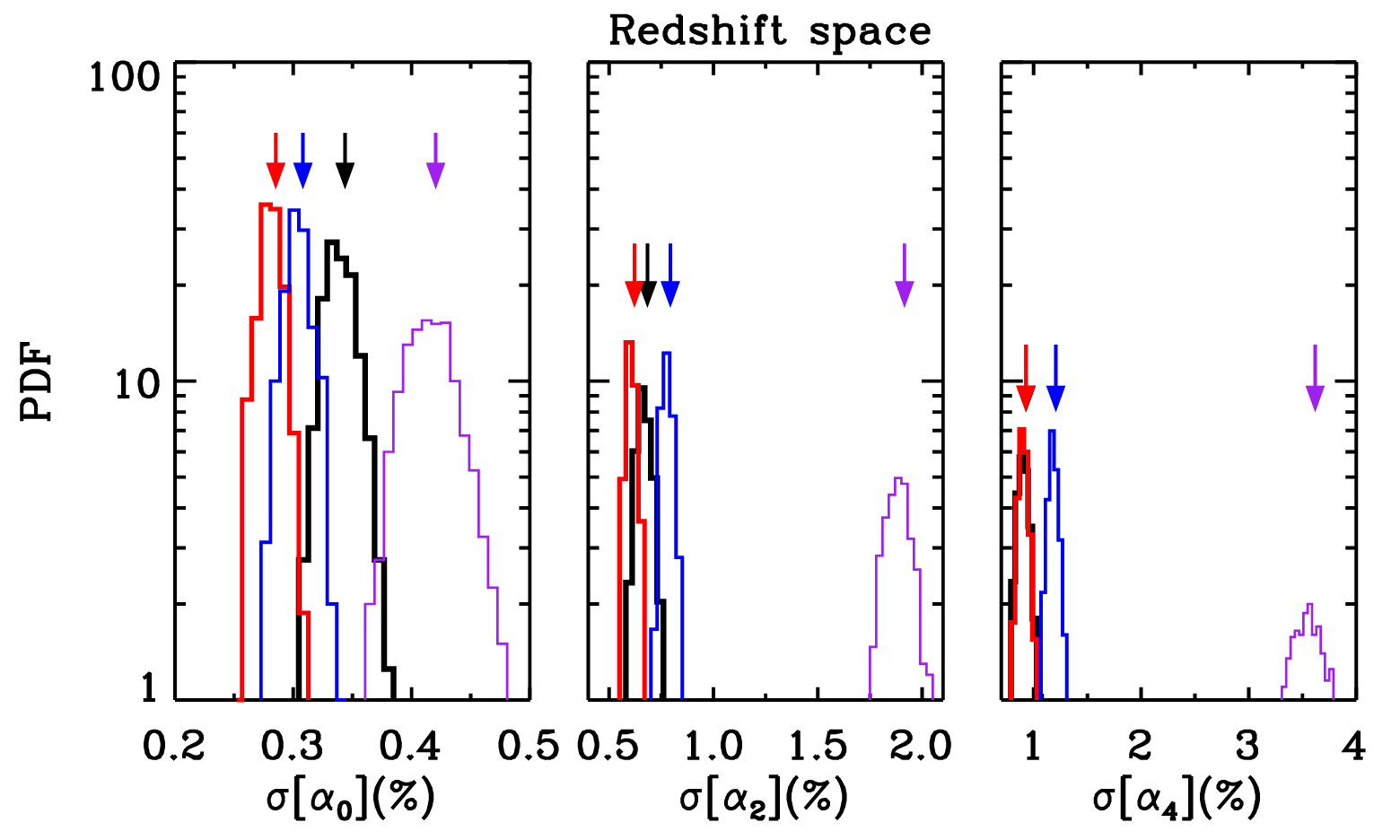}
\begin{center}
\end{center}
\caption{
\label{fig:8}
Distribution of $\alpha_\ell$ (left panels) and their uncertainties (right
panels) from MCMC analysis of $1\,000$ independent catalogues of DM particles
with $n=10^{-2}\,h^3\,\Mpc^{-3}$ and different Gaussian photo-$z$ errors. The
precision measuring $\alpha_\ell$ is computed after marginalising over the other
parameters in the model. The top and bottom rows show the results in real and
redshift space, respectively. Arrows point to the mean of each distribution. For
samples with sub-percent photo-$z$ errors, the value of $\alpha_\ell$ is
compatible with the fiducial cosmology at the $1\sigma$ level
($\alpha_\ell-1=0$). Therefore, sub-percent photo-$z$ errors do not introduce an
additional shift in the position of BAO. As we can see in the right panels, in
real space the precision measuring $\alpha_\ell$ decreases with the size of
photo-$z$ errors. Nevertheless, in redshift space samples with small photo-$z$
errors measure $\alpha_0$ with more precision than samples with no errors. This
is because the suppression of BAO in power spectrum moments decreases with the
size of photo-$z$ errors (see \S\ref{sec:3a}).
}
\end{figure*}

\begin{table}
\begin{center}
\caption{Average results extracted from the MCMC analysis of $1\,000$
independent catalogues of DM particles with $n=10^{-2}h^3\Mpc^{-3}$. We show the
average value of $\alpha_\ell$, $\bar{\alpha}_\ell$, and the average uncertainty
measuring this parameter after marginalising over the other free parameters in
the model.}
\label{tab:5}
\begin{tabular}{cccc}
\hline
&$\bar{\alpha}_0-1\,(\%)$
&$\bar{\alpha}_2-1\,(\%)$
&$\bar{\alpha}_4-1\,(\%)$\\\hline
\multicolumn{4}{c}{Real space}\\ \hline
0   & $0.20 \pm 0.22$ &$0.23 \pm 0.40$&$0.24 \pm 0.52$\\
0.3 & $0.19 \pm 0.25$ &$0.20 \pm 0.50$&$0.44 \pm 0.78$\\
0.5 & $0.14 \pm 0.30$ &$0.20 \pm 0.78$&$0.11 \pm 1.47$\\
1.0 & $0.30 \pm 0.40$ &$0.29 \pm 1.62$&$2.10 \pm 4.81$\\\hline
\multicolumn{4}{c}{Redshift space}\\ \hline
0   & $0.32 \pm 0.34$ &$0.40 \pm 0.68$&$0.44 \pm 0.93$\\
0.3 & $0.28 \pm 0.28$ &$0.34 \pm 0.62$&$0.33 \pm 0.93$\\
0.5 & $0.09 \pm 0.31$ &$0.23 \pm 0.79$&$0.18 \pm 1.20$\\
1.0 & $0.09 \pm 0.42$ &$0.38 \pm 1.92$&$-1.73 \pm 3.62$\\\hline
\end{tabular}
\end{center}
\end{table}

\subsection{Effect of redshift errors} \label{sec:5b}

In this section we apply our BAO analysis procedure to power spectrum moments of
samples with $n=10^{-2}h^3\Mpc^{-3}$ and Gaussian photo-$z$ errors of different 
sizes.

We start by studying the shift in the stretch parameter with respect to its
fiducial value, $\alpha_\ell-1$, from the BAO analysis of real- and
redshift-space power spectrum moments of $1\,000$ samples from the COLA
ensemble. In the left and right panels of Fig.~\ref{fig:8} we present the
distribution of these parameters and their uncertainties, respectively, from the
analysis of samples with different photo-$z$ errors. Arrows point to the mean of
each distribution, and we gather these values in Table~\ref{tab:5}. As we can
see in the left panels, the average value of $\alpha_\ell-1$ is compatible with
zero at the $1\sigma$ level for all samples. This implies that {\it for samples
with sub-percent photo-$z$ errors our estimator is unbiased relative to the case
with no errors}. However, for larger photo-$z$ errors this is no longer correct,
as we can see for $\alpha_4$ and $\sigma_z=1.0\,\%$. This is because our
procedure to calculate a smooth version of power spectrum moments does not work
well if the amplitude of the moment is of the order of the shot noise level.

In real and redshift space the stretch parameter presents a small and positive
shift, which is caused by the non-linear evolution of the matter density field
\citep[e.g.,][]{Angulo2008, Crocce2008, Smith2008, Padmanabhan2009}. We also
find that this shift slightly decreases with the size of photo-$z$ errors. This
is because photo-$z$ errors strongly suppress $k$-modes along the LOS on small
scales, which are the $k$-modes that present a greater coupling due to
non-linearities and RSD, and thus reduce the effective shift in the volume-averaged
stretch parameters. Whereas the shifts that we find are statistically
significant for the volume of the COLA ensemble, $27\,000\,h^{-3}\,\Gpc^3$, for
a single simulation they are compatible with zero to within $1\sigma$.
Furthermore, these shifts can in principle be approximately corrected for via
reconstruction algorithms\footnote{We note that this procedure has never been
applied to reconstruct the 3D density field of galaxy samples with photo-$z$
errors.} \citep[e.g.,][]{Eisenstein2007, Schmittfull2015} or by recalibrating the
estimator employed.

The top-right panel of Fig.~\ref{fig:8} displays the uncertainty in the
stretch parameter measured from samples in real space. As we expect, the
precision measuring $\alpha^r_\ell$ decreases with the size of photo-$z$ errors.
This is the consequence of photo-$z$ errors suppressing the clustering on small
scales, which increases the relative contribution of shot noise to the variance,
and thus reduces the ${\rm SNR}$.

In redshift space the precision measuring $\alpha_\ell$ is a combination of the
effect explained in the previous paragraph and the fact that photo-$z$ errors
reduce the overall suppression of BAO (see Fig.~\ref{fig:5}). As explained in
\S\ref{sec:3a}, they reduce the weight of $k$-modes parallel to the LOS when
doing the angular average of the 3D power spectrum, where these modes are
noisier than the perpendicular ones owing to RSD. In the bottom-right panel we
can see that the balance between both effects causes $\sigma[\alpha_\ell]$ not
to be a monotonic function of the size of photo-$z$ errors. We find that
$\alpha_0$ can be measured with more precision from samples with small photo-$z$
errors ($\sigma_z\leq 0.5\,\%$) than from samples with no errors. Nevertheless,
if we increase the size of photo-$z$ errors this is no longer true, for instance
$\sigma[\alpha_0]$ for samples with $\sigma_z=1.0\,\%$ is greater than for
samples with no photo-$z$ errors. For higher order moments the results are different
because the range of scales for which the variance is not dominated by shot
noise is smaller. However, for samples with $\sigma_z= 0.3\,\%$ we still find
that $\alpha_2$ and $\alpha_4$ can be measured with more precision or the same
precision, respectively, as from samples with no photo-$z$ errors.

A by-product of the MCMC analysis of power spectrum moments is the value of the
parameters $(\sigma_\perp,f)$, which encode the suppression of the moments
perpendicular and parallel to the LOS ($\sigma_\parallel=f\sigma_\perp$),
respectively. We find that their average value and uncertainty are approximately
the same independently of the size of photo-$z$ errors. For samples with no
photo-$z$ errors they are $\sigma_\perp=4.02\pm 0.13$ and $f=1.06 \pm 0.05$ in
real space, and $\sigma_\perp=4.29\pm 0.38$ and $f=1.68\pm 0.25$ in redshift
space. As expected, RSD increase the suppression of $k$-modes parallel to the
LOS: for the COLA ensemble at $z=1$ this suppression is $68\,\%$ greater than
for $k$-modes perpendicular to the LOS. This is the main reason behind getting
more precise results in $\alpha_\ell$ from samples in redshift space with
sub-percent photo-$z$ errors for which shot noise is not relevant on BAO scales.


\begin{figure}
\begin{center}
\includegraphics[width=0.475\textwidth]{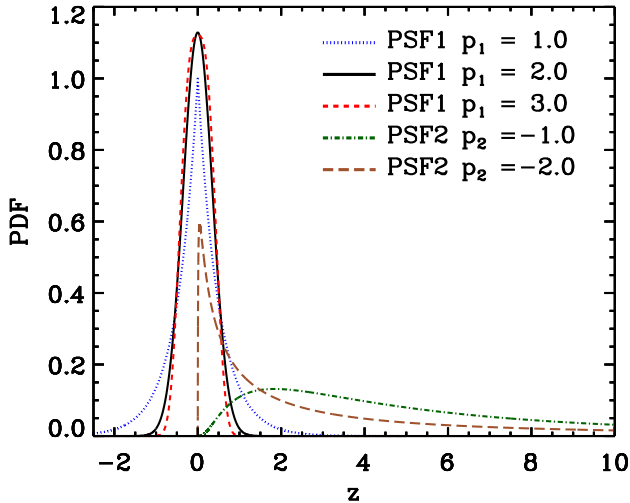}
\end{center}
\caption{
\label{fig:10}
Distributions from which photo-$z$ errors are drawn in \S\ref{sec:5c}. The
parameter $p_1$ controls the excess kurtoris of distributions from {\small PDF}1
and $p_2$ the skewness and excess kurtosis of that from {\small PDF}2.
All distributions from {\small PDF}1 are symmetric (zero skewness), where the
ones with $p_1=2$, $p_1>2$, and $p_1<2$ are Gaussians, boxier than Gaussians, and
Gaussians with extended wings like Lorentzians, respectively. The distributions
from {\small PDF}2 are positively skewed, and their skewness and excess kurtosis
grow with $p_2$.
}
\end{figure}


\begin{figure}
\begin{center}
\includegraphics[width=0.475\textwidth]{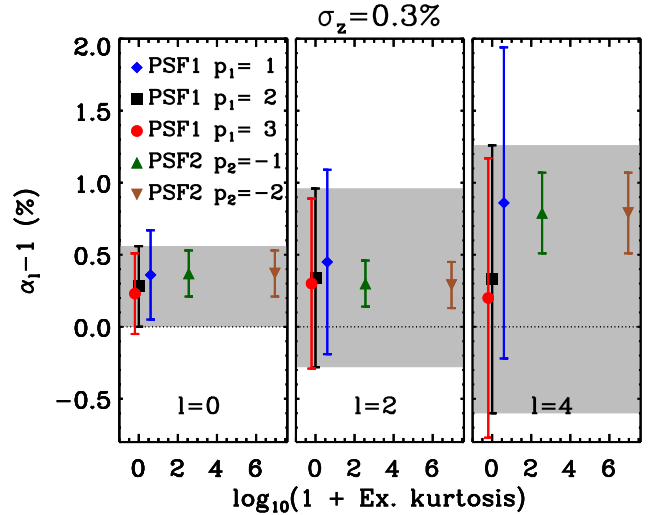}
\end{center}
\caption{
\label{fig:11}
Shift and uncertainty in $\alpha_0$, $\alpha_2$, and $\alpha_4$ from BAO
analysis of 100 samples from the COLA ensemble with $n=10^{-2}h^3\Mpc^{-3}$ and
photo-$z$ errors drawn from different PDFs, where in the analysis we assume that
they are drawn from a Gaussian PDF. The grey coloured areas enclose the
$1\,\sigma$ confidence region from the analysis of samples with Gaussian
photo-$z$ errors. The employed PDFs are shown in Fig.~\ref{fig:10}. Even for
photo-$z$ errors drawn from PDFs with large excess kurtosis and skewness, the
results are compatible with the Gaussian case at the $1\,\sigma$ level.}
\end{figure}

\subsection{Effect of different PDFs for photometric redshift errors} 
\label{sec:5c}

In general, photo-$z$ errors do not follow a Gaussian PDF. For instance, the
comparison between photometric and spectroscopic redshifts in the COSMOS survey
shows that the PDF of photo-$z$ errors is well described by a Lorentzian variate
\citep{Ilbert2009}. Additionally, for low redshift galaxies the PDF of photo-$z$
errors usually shows a tail towards higher redshifts, which is a natural
consequence of imposing $z > 0$ in an otherwise symmetric PDF. In this section
we investigate whether photo-$z$ drawn from non-Gaussian PDFs may bias the
results from BAO analyses when assuming that they are drawn from a Gaussian
PDF.

We will consider two families of functional forms for photo-$z$ errors:

\begin{itemize}
\item[i)] {\small PDF}1:
\begin{equation}
\Pr[\delta r_z]\text{d}z=\frac{1}{2\,\Delta\,\Gamma\left(1+\frac{1}{p_1}\right)}
\exp\left(-\left|\frac{z}{\Delta}\right|^{p_1}\right)\text{d}z,
\end{equation}

\item[ii)] {\small PDF}2: 
\begin{equation}
\Pr[\delta r_z]\text{d}z = \frac{-1}{p_2\,z\sqrt{2\pi}}
\exp\left[-\frac{1}{2p_2^2}\text{ln}^2
\left(-\frac{p_2\,z}{\Delta}\right)\right] \text{d}z,
\end{equation}
\end{itemize}

\noindent where $\Gamma$ is the Gamma function, $\Delta$ controls the width of
the distributions, $p_1$ the excess kurtosis for the family {\small PDF}1, and
$p_2$ the skewness and excess kurtosis for the family {\small PDF}2. The
distributions from {\small PDF}1 are symmetric (zero skewness) and show
different levels of excess kurtosis. We find that for $p_1=2$ the distributions
of this family are Gaussians, for $p_1<2$ show extended wings like a Lorentzian,
and for $p_1>2$ are boxier than a Gaussian. The distributions from {\small PDF}2
are asymmetric and not centred in zero. We find one famous distribution of this
family for $\Delta=p_2$ and $p_2<0$, the log-normal distribution. 

In Fig.~\ref{fig:10} we display the PDF of distributions that we use to
introduce photo-$z$ errors in the COLA ensemble. We note that the ones with
$p_2\neq 0$ are in general more extreme than the PDF of photo-$z$ errors from
real data. Therefore, if we were to find that the results from BAO analyses
assuming a Gaussian distribution are the same for all these distributions, we
could conclude that BAO analyses are insensitive to the shape of the PDF of
photo-$z$ errors.

In Fig.~\ref{fig:11} we present the value of $\alpha_0$, $\alpha_2$, and
$\alpha_4$ extracted from the BAO analysis of the average moments of 100 samples
from the COLA ensemble after assuming a Gaussian PDF in the analysis. The number
density of these samples is $n=10^{-2}h^3\Mpc^{-3}$, their photo-$z$ errors are
drawn from the distributions displayed in Fig.~\ref{fig:10}, and the difference
between the 84th and 16th percentiles of those distributions is set to be
$\sigma_z=0.3\,\%$. The grey coloured regions indicate the $1\,\sigma$
confidence region for a Gaussian PDF, and the error bars for the other
distributions. For extreme PDFs, this could in principle introduce systematic
errors in the estimation of $\alpha_\ell$. In practice we can see that even
considering extreme PDFs and assuming a Gaussian PDF in the BAO analysis, the
results are compatible to within $1\sigma$. In addition, the shift in the
stretch parameter is largely insensitive to the actual shape of the PDF.

In this section we have disregarded the possibility of interlopers - objects
systematically assigned to incorrect redshifts. This may happen to unobscured
quasars and star-forming galaxies in medium- and narrow-band surveys, as pairs
of emission lines at different redshifts may fall in the same filters, which is
translated into a redshift PDF with two or more peaks \citep[see, e.g., fig. 4
of][]{chavesmontero17}. If the percentage of interlopers is very small, at first
order their net effect is to increase the shot noise level as they are
uncorrelated with the main sample. They can be accounted for by artificially
increasing the shot noise level. Nonetheless, if their amount is significant
with respect to the main sample, they introduce anisotropies in the galaxy
clustering\citep{Lidz2016}, where these anisotropies open the possibility of
using AP-type tests \citep{Alcock1979} to correct for them.

We note that all expressions in this work are given for arbitrary PDFs.
Therefore, if it is possible know the PDF of photo-$z$ errors, its actual shape
can be used in BAO analyses.


\begin{figure}
\begin{center}
\includegraphics[width=0.475\textwidth]{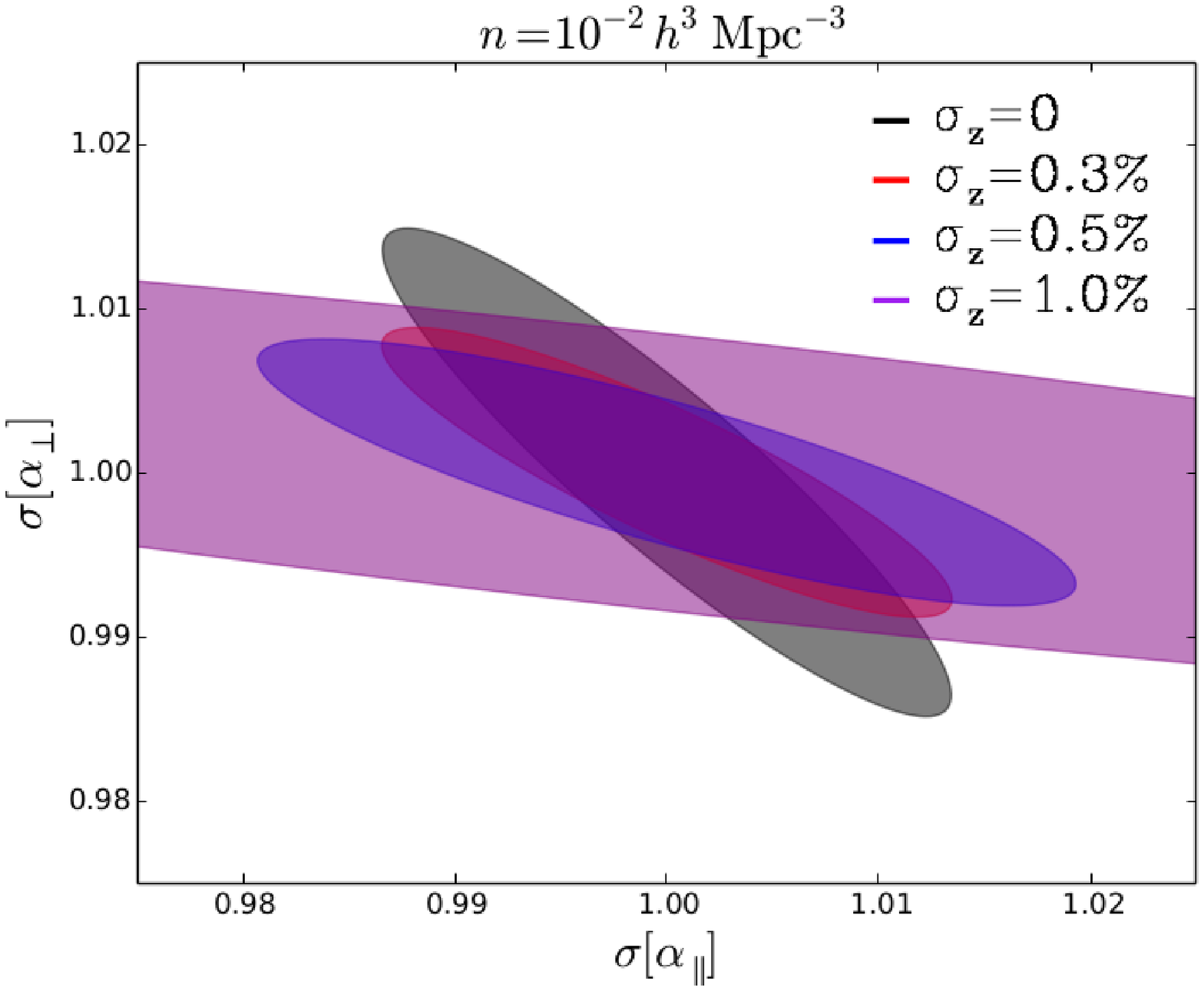}

\includegraphics[width=0.475\textwidth]{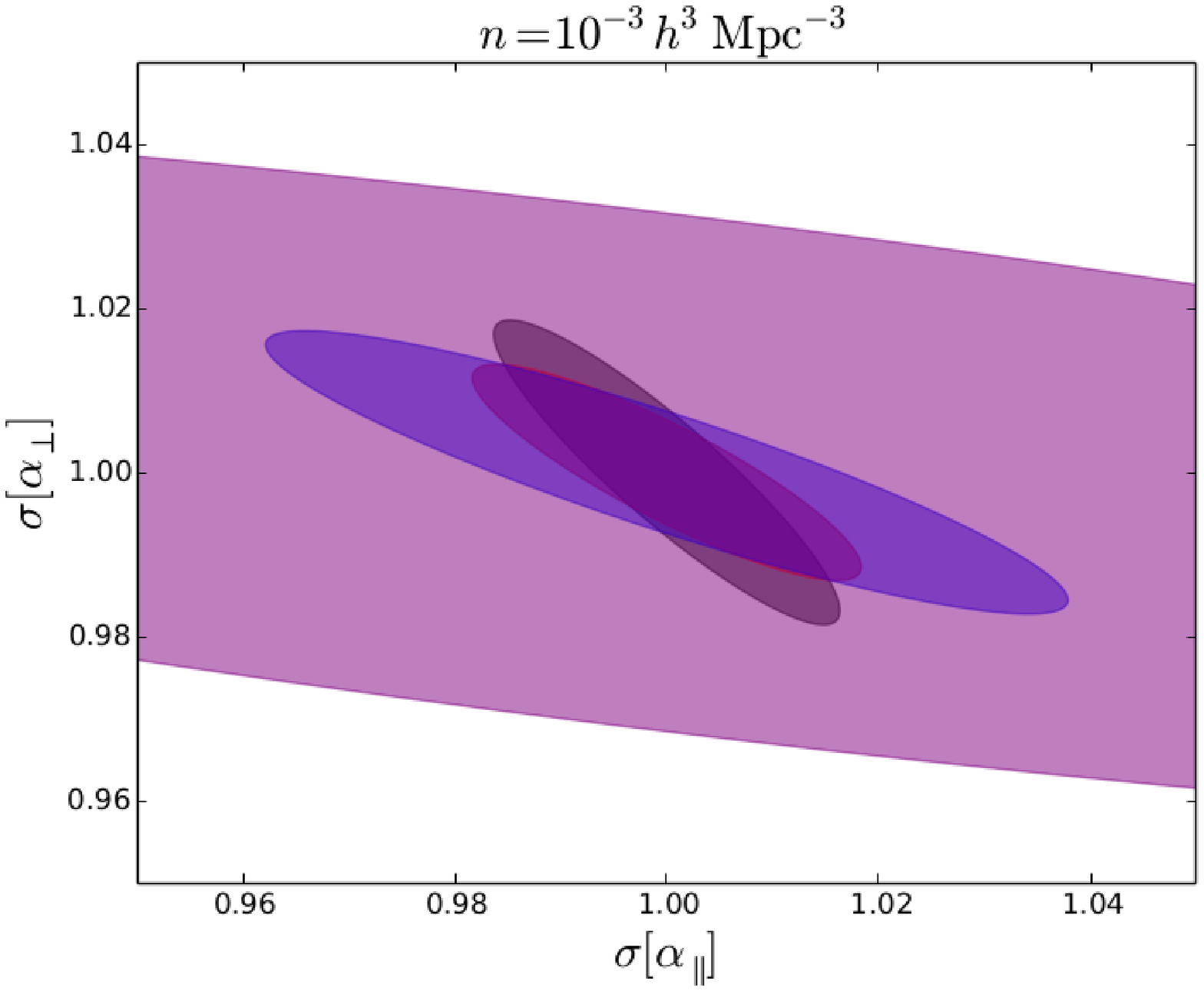}
\end{center}
\caption{
\label{fig:12}
Uncertainty in the parallel and perpendicular components of the stretch
parameter for samples with $n=10^{-2}$ and $10^{-3}\,h^3\,\Mpc^{-3}$ (top and
bottom panel, respectively), where they control the precision measuring $H$ and
$D_A$, respectively. As we can see, the uncertainty in $H$ after marginalising
over $D_A$ grows with the size of photo-$z$ errors, as the error ellipses rotate
in the anticlockwise direction. However, the FoM (inverse of the ellipse's area)
does not monotonically decreases with $\sigma_z$, which means that the
combination of $H$ and $D_A$ is not always measured with more precision by
samples with no photo-$z$ errors from the analysis of power spectrum moments.
}
\end{figure}



\begin{figure*}
\begin{center}
\includegraphics[width=0.95\textwidth]{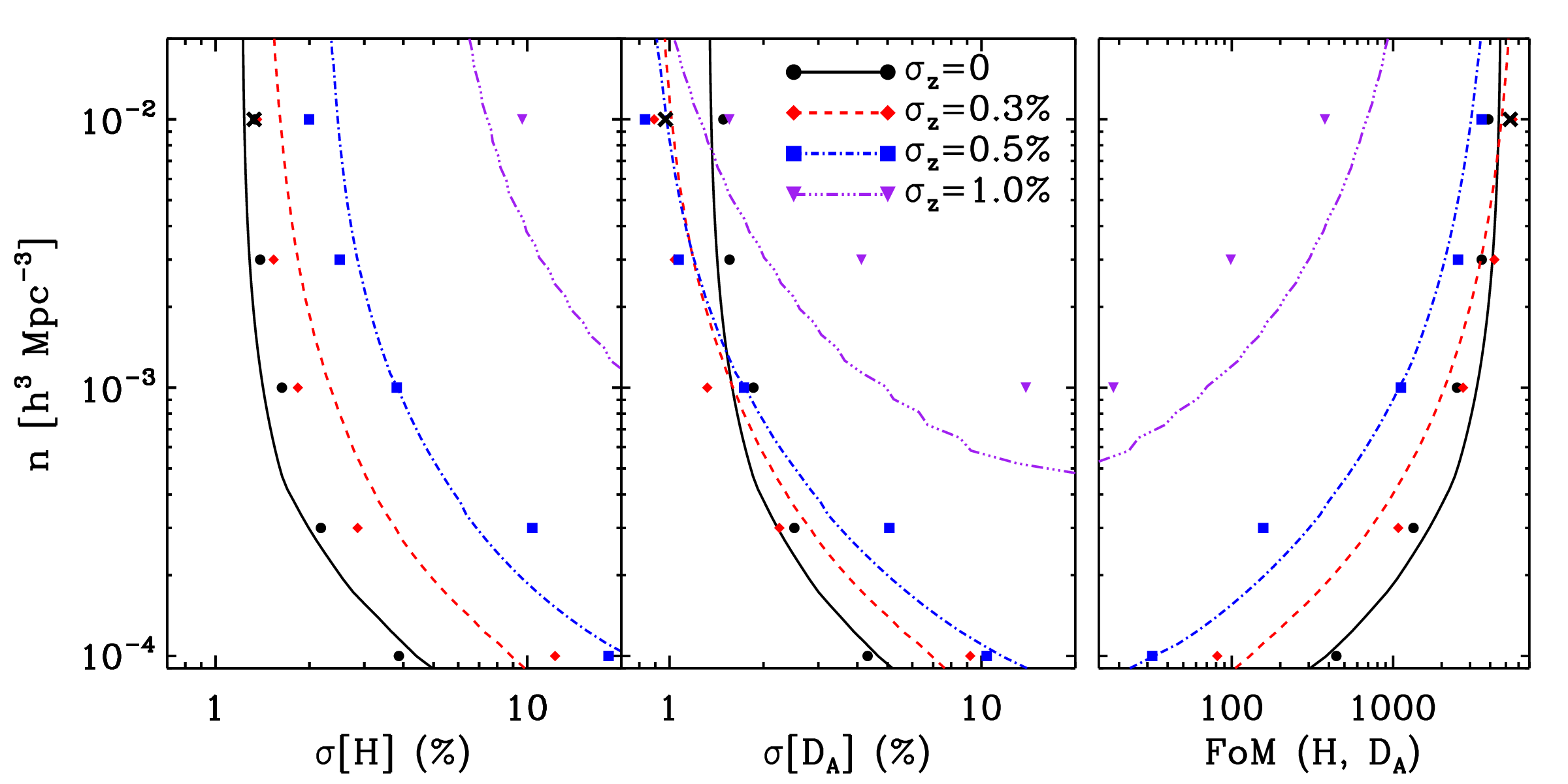}
\end{center}
\caption{
\label{fig:13}
Precision measuring the Hubble parameter (left panel), the angular diameter
distance (middle panel), and their FoM from the BAO analysis of samples with
different number densities and Gaussian photo-$z$ errors. Each symbols indicate
the average result from $1\,000$ samples from the COLA ensemble and the lines
show analytic predictions from Eq.~\ref{eq:intsneff} using $A_0=0.35$,
$A_2=0.50$, $A_4=0.55$, and $c=0.3$. The uncertainty in $H$ grows with the size
of photo-$z$ errors and by decreasing the number density. On the other hand, the
behaviour of $\sigma[D_A]$ is more complex, where samples with sub-percent
photo-$z$ errors have more precision measuring $D_A$ than samples with no
errors. This highlights that all the cosmological information encoded on BAO is
not recovered from the analysis of power spectrum moments. Black crosses
indicate the results after deconvolving the effect of RSD using
Eq.~\ref{eq:deconvolve} in samples with $10^{-2}h^3\Mpc^{-3}$ and no photo-$z$
errors. This simple algorithm increases by $54\,\%$ the precision measuring
$D_A$.
}
\end{figure*}


\subsection{Extracting cosmological information from BAO and impact of number
density} 
\label{sec:5d}

As explained in \S\ref{sec:5b}, the constraining power of BAO depends on the
scale at which the shot noise level starts dominating the amplitude of power
spectrum moments. In this section we address how the precision in measuring
cosmological parameters from BAO analyses ($H$ and $D_A$) depends on the number
density of the analysed sample. For this, we re-analyse our COLA samples
randomly diluted to have $\bar{n} = [10^{-2},\, 3\times10^{-3},
\,10^{-3}, \,3\times10^{-4}, \,10^{-4}]\,h^3\Mpc^{-3}$.

In the top and bottom panels of Fig.~\ref{fig:12} we present the average
precision in measuring $\sigma_{\perp}$ and $\sigma_{\parallel}$ from $1\,000$
COLA samples with $n=10^{-2}$ and $10^{-3}\,h^3\,\Mpc^{-3}$, respectively.
Contours enclose the $1\,\sigma$ confidence region for samples with different
Gaussian photo-$z$ errors, as stated in the legend. The uncertainty in
$\sigma_{\parallel}$ (after marginalising over $\sigma_{\perp}$) always
increases with the size of photo-$z$ errors, which is because the error ellipses
rotate anticlockwise as photo-$z$ errors grow. Nevertheless, the uncertainty in
$\sigma_{\perp}$ and the Figure-of-Merit (FoM) of this combination of parameters
(the inverse of the ellipse's area) do not monotonically grow with the size of
photo-$z$ errors. For the number densities studied here, we find that for
$\sigma_z=0.3\,\%$ the precision in measuring $D_A$ is greater than for samples
with $\sigma_z=0$, which highlights that the analysis of power spectrum moments
might not optimally extract the cosmological information encoded in BAO
perpendicular to the LOS. We further explore this in the next section.

To continue exploring the constraints on cosmological parameters as a function
of number density and photo-$z$ errors, in the left, middle, and right panels of
Fig.~\ref{fig:13} we display the precision in $H$, $D_A$, and the FoM of
both parameters, respectively. The uncertainty in each parameter is computed
after marginalising over the other. Symbols and lines indicate the results from
simulations and the analytic model introduced in Eq.~\ref{eq:intsneff},
respectively, where the free parameters of the model are fitted to reproduce the
results from simulations. Their value is $A_0=0.35$, $A_2=0.50$, $A_4=0.55$, and
$c=0.3$. As expected, the uncertainty in $H$ grows with the size of photo-$z$
errors and by decreasing the number density. However, it is important to notice
that the precision in $H$ is the same for samples with $\sigma_z=0.5\,\%$
and $n=10^{-3}\,h^3\,\Mpc^{-3}$ and as for samples no photo-$z$ errors and
$10^{-4}\,h^3\,\Mpc^{-3}$. As spectro-photometric surveys detect in general
fainter objects than spectroscopic surveys, future wide-field surveys with
dozens of photometric bands such as J-PAS will be competitive with spectroscopic
surveys measuring cosmological parameters from BAO analyses. 

The precision measuring $D_A$ and the FoM of $H$ and $D_A$ shows a non-monotonic
behaviour with $\sigma_z$ for samples with large number densities, whereas at
smaller number densities they are proportional to the size of photo-$z$ errors.
As we have commented above, we leave the discussion of this to the following
section.

Our analytic model, which only employs the real-space $l=0$ moment as input,
reasonably fits the results from simulations. Therefore, it can be used to make
forecasts for the precision measuring cosmological parameters from samples with
different number densities, linear biases, photo-$z$ errors, and cosmologies.
Furthermore, we can use this model to look for the best sample to constrain
cosmology. All the above considerations should be taken into account for the
optimal design of future galaxy surveys. For instance, the photo-$z$ errors for
a given galaxy sample might not only depend on the hardware employed, but also
on the intrinsic properties of galaxies (e.g. brighter objects having more
accurate redshift estimates). In such case, the sample that delivers the
strongest constraints on cosmological parameters is not necessarily the one with
the smallest photo-$z$ errors.


\begin{figure}
\begin{center}
\includegraphics[width=0.475\textwidth]{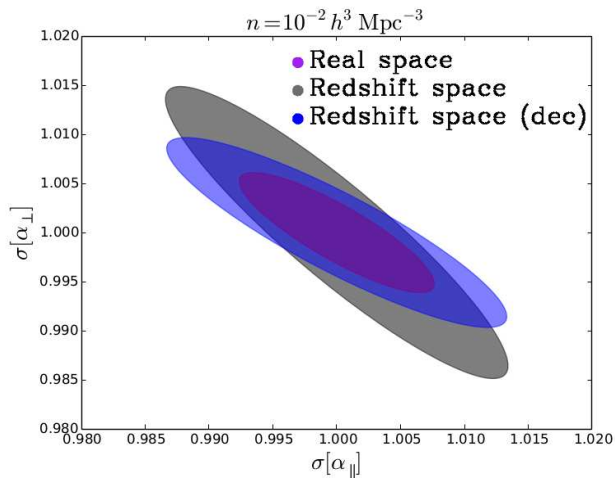}
\end{center}
\caption{
\label{fig:14}
Same as Fig.~\ref{fig:12} for samples with $n=10^{-2}\,h^3\,\Mpc^{-3}$ and no
photo-$z$ errors in real space, redshift space, and redshift space after
deconvolving the effect of RSD (in purple, grey, and blue, respectively). The
strongest constraints on $H$ ($\sigma[\alpha_\parallel]$) come from real-space
moments because RSD further suppress BAO along the LOS. The constraints on $D_A$
($\sigma[\alpha_\perp]$) should be approximately the same in real- and
redshift-space; however, they precision measuring $D_A$ is 2.4 times smaller in
redshift space. As we can see, this is partially corrected by the deconvolution
procedure introduced in \S\ref{sec:5e}.
}
\end{figure}


\subsection{Loss of transverse information from the analysis of power spectrum
moments}
\label{sec:5e}

As we showed in the previous section, for large number densities $D_A$ is
measured with more precision from samples with sub-percent photo-$z$ errors than
from samples with no errors. We obtain the same results from simulations and
from the analytic model introduced in Eq.~\ref{eq:intsneff}. As the introduction
of photo-$z$ errors cannot increase the amount of cosmological information, this
highlights that the analysis of power spectrum moments does not extract all the
cosmological information encoded in BAO. Therefore, it is worth to follow other
approaches such as the analysis of the full anisotropic power spectrum
\citep[e.g.,][]{Ballinger96}.

This reduction of the cosmological information available from BAO analyses only
appears in redshift space. RSD suppress more strongly parallel $k$-modes, and
when we take the angular average of the power spectrum to compute its moments,
we treat in the same way all $k$-modes, even when the ones parallel to the LOS
are noisier. Consequently, the resulting power spectrum obtained after averaging
over all $k$-modes in a $k$-bin is noisier than if only perpendicular $k$-modes
are considered. For samples with photo-$z$ errors this is not the case, as they
reduce the weight of parallel $k$-modes during the angular average, and thus the
uncertainty after angular averaging over all $k$-modes is approximately the same
as for perpendicular $k$ modes (c.f. \S\ref{sec:3a}).

To probe that the angular average of $k$-modes with different uncertainties
causes $D_A$ to be measured with smaller precision, we deconvolve the effect of
RSD from power spectrum moments:

\begin{equation}
\label{eq:deconvolve}
\tilde{P}(k)= \left\langle \frac{n\hat{P}(k,\mu)-1}{n\f^2(k,\mu)} \right\rangle_{\hat{\vk}}.
\end{equation}

In Fig.~\ref{fig:14} we show the average uncertainty in cosmological
parameters computed from $1\,000$ samples of the COLA ensemble with
$n=10^{-2}\,h^3\,\Mpc^{-3}$ and no photo-$z$ errors. The purple, grey, and blue
ellipses indicate the results in real space, redshift space, and redshift space
after deconvolving the effect of RSD. Our deconvolution procedure increases the
precision measuring $D_A$ and the FoM of $H$ and $D_A$ by $54\,\%$ and $37\,\%$,
respectively, and, as expected, it does not reduce $\sigma[H]$. Nevertheless,
our naive approach does not totally correct the effect of RSD, as real-space
power spectrum moments measure $H$ and $D_A$ with greater precision.

We note that Eq.~\ref{eq:deconvolve} cannot be applied to samples with photo-$z$
errors on intermediate and small scales. This is because they strongly suppress
$k$-modes with $\mu\simeq 1$, which causes them to be completely dominated by
shot noise, and our shot noise subtraction is not accurate enough in this
regime. Nonetheless, a correct characterisation of shot noise in this regime
could lead to a joint deconvolution of RSD and photo-$z$ errors. We note that
there are other approaches in the literature to reduce the impact of photo-$z$
errors on the power spectrum \citep[e.g.,][]{McQuinn2013}.


\begin{figure}
\begin{center}
\includegraphics[width=0.475\textwidth]{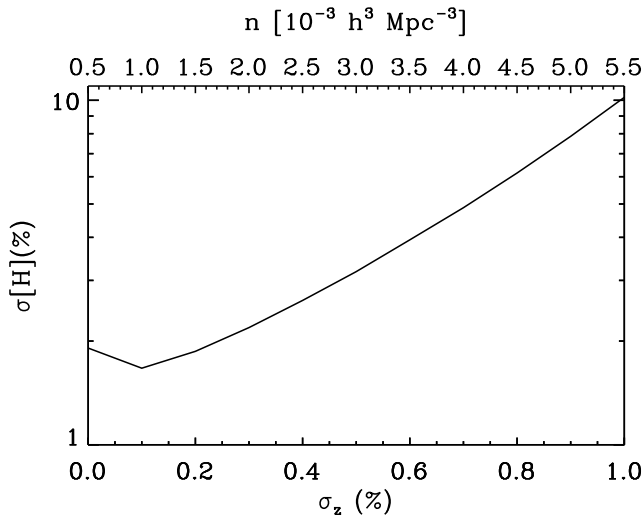}
\end{center}
\caption{
\label{fig:15}
Forecasts for the precision measuring $H$ in spectro-photometric galaxy
surveys at $z=1$. The results are computed assuming a comoving volume of
$V=78.7\,\Gpc^3$ and that the number density of galaxies linearly scales with
$\sigma_z$. As we can see, galaxy samples with $\sigma_z\leq 0.3\,\%$ measure the
Hubble parameter with approximately the same precision.
}
\end{figure}

\section{Forecasts for future galaxy surveys} 
\label{sec:6o}

In \S\ref{sec:3c} we introduced an analytic expression to compute the precision
measuring cosmological parameters from BAO analyses, and in the previous section
we showed that this model approximately reproduces the results from numerical
simulations. In this section we use this expression to forecast the precision in $H$
from future spectro-photometric surveys at $z=1$. We note that the results of this
section are illustrative.

In what follows we will assume that the number density of galaxies linearly
scale with the size of photo-$z$ errors \citep[see table 8 of][]{benitez14} and
that the analysed volume is $V=78.7\,\Gpc^3$, i.e. the same volume as each COLA
simulation. In particular, we take the relation between number density and
photo-$z$ errors to be $n=5(1+10^3\sigma_z)10^{-4}\,h^3\,\Mpc^{-3}$. In
Fig.~\ref{fig:15} we display the precision measuring $H$ after marginalising
over $D_A$ from galaxy samples with $b=2$ and different photo-$z$ errors. For
samples with $\sigma_z\leq 0.3\,\%$, the precision measuring $H$ is
approximately the same. On the other hand, for $\sigma_z> 0.3\,\%$ the
uncertainty in $H$ is rapidly increased. This encourages the employment of
spectroscopic and spectro-photometric surveys with $\sigma_z< 0.5\,\%$ to study
the expansion history of the Universe.

In summary, to design and fully exploit galaxy surveys that employ noisy
estimators to compute redshifts, it is necessary to carefully select the
properties of the target galaxy sample.

\section{Conclusions} \label{sec:7o}

The next generation of galaxy surveys will dramatically increase the precision
of measurements for the expansion and growth history of the Universe. Some of
these surveys will observe large areas of the sky with linear variable filters
or dozens of narrow-bands, providing a low-resolution spectra for every region 
of the sky. In addition, they will measure the redshift of millions of 
galaxies with sub-percent accuracy, offering a promising way of constraining
cosmological parameters. Nevertheless, to fully exploit this new kind of data
it is necessary to fully characterise the effect of photo-$z$ errors on
cosmological observables.

In this work we presented a detailed study of the impact of sub-percent
photo-$z$ errors on the clustering of galaxies in Fourier space, with an
emphasis on the BAO signal. We derived analytic expressions for how photo-$z$
errors modify power spectrum moments, their variances, and the smearing of
BAO, which we compared with the results from $1\,000$ $N$-body simulations.

Our main findings can be summarised as follows:

\begin{itemize}

\item In real space photo-$z$ errors suppress power spectrum moments on
intermediate and small scales. This increases the interval of scales dominated
by shot noise, which reduces the range of scales available for BAO analyses.
There is an additional effect in redshift space: the suppression of
angular-averaged BAO wiggles gets weaker with increasing photo-$z$ errors. This
is because photo-$z$ errors reduce the weight of LOS $k$-modes in computing
power spectrum moments which have more diluted BAO signal due to nonlinear RSD.

\item We derived how the cosmological information encoded in BAO depends on the
properties of the galaxy sample studied. We showed that small-scale RSD and/or
photo-$z$ errors induce a scale-dependence on this information, where the
dependence on the Hubble parameter (angular diameter distance) decreases
(increases) with the size of photo-$z$ errors.

\item Based on these findings, we built a model for extracting cosmological
information from the analysis of power spectrum moments. Then, we applied it to
simulated galaxy catalogues with different number densities and photo-$z$
errors. We found that photo-$z$ errors do not introduce an additional shift in
the position of the BAO scale with respect to the no photo-$z$ error case.
Therefore, they do not bias the cosmological information encoded in BAO. In
addition, we found that assuming that photo-$z$ errors are Gaussian in BAO
analyses, even when they are drawn from PDFs with large excess kurtosis and
skewness, does not bias the results.

\item In \S\ref{sec:5d} we analysed the precision measuring the Hubble parameter
and the angular diameter distance from samples with different number densities
and photo-$z$ errors. We found that for the same number density, the uncertainty
in measuring $H$ decreases with the size of photo-$z$ errors. Nevertheless, it
is still possible to measure $H$ with the same (or more) precision from samples
with sub-percent photo-$z$ errors as from samples with no errors if the number
density of the first is increased. Finally, we also found that the analysis of
power spectrum moments artificially decreases the precision in measuring $D_A$.
We suggest to analyse the 2D power spectrum in future studies.

\end{itemize}

Our results encourage the measurement of cosmological parameters from
spectro-photometric surveys, as in general they are deeper than spectroscopic
surveys for the same integration time. In \S\ref{sec:6o} we forecast the
precision in measuring the Hubble parameter from BAO analyses assuming that the
number density of galaxies linearly scales with the size photo-$z$ errors.
Roughly, we found the same results for galaxy samples with redshift
uncertainties smaller than $\sigma_z=0.4\,\%$. This means that galaxy surveys
with sub-percent photo-$z$ errors could set constraints on the dark energy
equation of state as precise as spectroscopic surveys.

Along this paper we put a focus on extracting cosmological information from
galaxy samples with sub-percent photo-$z$ errors. Recently, \citet{Ross17}
conducted a similar investigation in configuration space for samples with
photo-$z$ errors of a few percent, finding that for those the BAO feature mostly
constrains $D_A(z)$ and that the projected correlation function is enough for
extracting all cosmological information. Their findings agree with ours for
samples with $\sigma_z = 1\,\%$, as we can see in Figs. \ref{fig:12} and
\ref{fig:13}. Nevertheless, as we show along this work, for samples with smaller
photo-$z$ errors $H(z)$ can be constrained from the 3D galaxy clustering.

Finally, our paper highlights that photo-$z$ errors substantially increase the
complexity of the extraction of cosmological information from BAO analyses.
Therefore, it is crucial a thorough understanding and modelling of photo-$z$
errors in galaxy clustering. We hope our work to have clarified some the most
important aspects of this issue, and that it will help in the cosmological
analysis of future spectro-photometric surveys.

\section*{Acknowledgements}

We thank the anonymous referee for the thorough review, insightful comments, and
positive suggestions. We acknowledge discussions with Raul Abramo, Andreu
Font-Ribera, Licia Verde, and Carlos L\'opez-Sanjuan. Argonne National
Laboratory's work was supported by the U.S. Department of Energy, Office of
Science, Office of Nuclear Physics, under contract DE-AC02-06CH11357. The
authors acknowledge support from the Spanish Ministry of Economy and
Competitiveness (MINECO) through the project AYA2015-66211-C2-2. JCM
acknowledges support from the Fundaci\'on Bancaria Ibercaja for developing this
research. REA acknowledges support from the European Research Council through
grant number ERC-StG/716151. CHM acknowledges support from the Ramon y Cajal
Fellow Program of the Spanish MINECO. This project has received funding from the
European Union’s Horizon 2020 Research and Innovation Programme under the Marie
Sklodowska-Curie grant agreement No 734374.

\bibliographystyle{mn2e}
\bibliography{database}

\begin{thebibliography}{57}
\expandafter\ifx\csname natexlab\endcsname\relax\def\natexlab#1{#1}\fi

\bibitem[{{Abramowitz} \& {Stegun}(1972)}]{Abramowitz72}
{Abramowitz} M., {Stegun} I.~A., 1972, {Handbook of Mathematical Functions}

\bibitem[{{Alcock} \& {Paczynski}(1979)}]{Alcock1979}
{Alcock} C., {Paczynski} B., 1979, \nat, 281, 358

\bibitem[{{Angulo} {et~al}\mbox{.}(2008){Angulo}, {Baugh}, {Frenk}, \&
  {Lacey}}]{Angulo2008}
{Angulo} R.~E., {Baugh} C.~M., {Frenk} C.~S., {Lacey} C.~G., 2008, \mnras, 383,
  755

\bibitem[{{Angulo} \& {Pontzen}(2016)}]{AnguloPontzen2016}
{Angulo} R.~E., {Pontzen} A., 2016, \mnras, 462, L1

\bibitem[{{Angulo} {et~al}\mbox{.}(2012){Angulo}, {Springel}, {White},
  {Jenkins}, {Baugh}, \& {Frenk}}]{Angulo2012}
{Angulo} R.~E., {Springel} V., {White} S.~D.~M., {Jenkins} A., {Baugh} C.~M.,
  {Frenk} C.~S., 2012, \mnras, 426, 2046

\bibitem[{{Angulo} {et~al}\mbox{.}(2014){Angulo}, {White}, {Springel}, \&
  {Henriques}}]{Angulo2014}
{Angulo} R.~E., {White} S.~D.~M., {Springel} V., {Henriques} B., 2014, \mnras,
  442, 2131

\bibitem[{{Ballinger} {et~al}\mbox{.}(1996){Ballinger}, {Peacock}, \&
  {Heavens}}]{Ballinger96}
{Ballinger} W.~E., {Peacock} J.~A., {Heavens} A.~F., 1996, \mnras, 282, 877

\bibitem[{{Benitez} {et~al}\mbox{.}(2014){Benitez}, {Dupke}, {Moles}, {Sodre},
  {Cenarro}, {Marin-Franch}, {Taylor}, {Cristobal}, {Fernandez-Soto}, {Mendes
  de Oliveira}, {Cepa-Nogue}, {Abramo}, {Alcaniz}, {Overzier},
  {Hernandez-Monteagudo}, {Alfaro}, {Kanaan}, {Carvano}, {Reis}, {Martinez
  Gonzalez}, {Ascaso}, {Ballesteros}, {Xavier}, {Varela}, {Ederoclite},
  {Vazquez Ramio}, {Broadhurst}, {Cypriano}, {Angulo}, {Diego}, {Zandivarez},
  {Diaz}, {Melchior}, {Umetsu}, {Spinelli}, {Zitrin}, {Coe}, {Yepes}, {Vielva},
  {Sahni}, {Marcos-Caballero}, {Shu Kitaura}, {Maroto}, {Masip}, {Tsujikawa},
  {Carneiro}, {Gonzalez Nuevo}, {Carvalho}, {Reboucas}, {Carvalho}, {Abdalla},
  {Bernui}, {Pigozzo}, {Ferreira}, {Chandrachani Devi}, {Bengaly}, {Campista},
  {Amorim}, {Asari}, {Bongiovanni}, {Bonoli}, {Bruzual}, {Cardiel}, {Cava},
  {Cid Fernandes}, {Coelho}, {Cortesi}, {Delgado}, {Diaz Garcia}, {Espinosa},
  {Galliano}, {Gonzalez-Serrano}, {Falcon-Barroso}, {Fritz}, {Fernandes},
  {Gorgas}, {Hoyos}, {Jimenez-Teja}, {Lopez-Aguerri}, {Lopez-San Juan},
  {Mateus}, {Molino}, {Novais}, {OMill}, {Oteo}, {Perez-Gonzalez}, {Poggianti},
  {Proctor}, {Ricciardelli}, {Sanchez-Blazquez}, {Storchi-Bergmann}, {Telles},
  {Schoennell}, {Trujillo}, {Vazdekis}, {Viironen}, {Daflon},
  {Aparicio-Villegas}, {Rocha}, {Ribeiro}, {Borges}, {Martins}, {Marcolino},
  {Martinez-Delgado}, {Perez-Torres}, {Siffert}, {Calvao}, {Sako}, {Kessler},
  {Alvarez-Candal}, {De Pra}, {Roig}, {Lazzaro}, {Gorosabel}, {Lopes de
  Oliveira}, {Lima-Neto}, {Irwin}, {Liu}, {Alvarez}, {Balmes}, {Chueca},
  {Costa-Duarte}, {da Costa}, {Dantas}, {Diaz}, {Fabregat}, {Ferrari},
  {Gavela}, {Gracia}, {Gruel}, {Gutierrez}, {Guzman}, {Hernandez-Fernandez},
  {Herranz}, {Hurtado-Gil}, {Jablonsky}, {Laporte}, {Le Tiran}, {Licandro},
  {Lima}, {Martin}, {Martinez}, {Montero}, {Penteado}, {Pereira}, {Peris},
  {Quilis}, {Sanchez-Portal}, {Soja}, {Solano}, {Torra}, \&
  {Valdivielso}}]{benitez14}
{Benitez} N. {et~al.}, 2014, ArXiv e-prints

\bibitem[{{Ben{\'{\i}}tez} {et~al}\mbox{.}(2009){Ben{\'{\i}}tez},
  {Gazta{\~n}aga}, {Miquel}, {Castander}, {Moles}, {Crocce},
  {Fern{\'a}ndez-Soto}, {Fosalba}, {Ballesteros}, {Campa}, {Cardiel-Sas},
  {Castilla}, {Crist{\'o}bal-Hornillos}, {Delfino}, {Fern{\'a}ndez},
  {Fern{\'a}ndez-Sopuerta}, {Garc{\'{\i}}a-Bellido}, {Lobo}, {Mart{\'{\i}}nez},
  {Ortiz}, {Pacheco}, {Paredes}, {Pons-Border{\'{\i}}a}, {S{\'a}nchez},
  {S{\'a}nchez}, {Varela}, \& {de Vicente}}]{Benitez2009}
{Ben{\'{\i}}tez} N. {et~al.}, 2009, \apj, 691, 241

\bibitem[{Bjerhammar(1951)}]{Bjerhammar51}
Bjerhammar A., 1951, Trans. Roy. Inst. Tech. Stockholm, 1951, 86 pp. (2 plates)

\bibitem[{{Blake} \& {Bridle}(2005)}]{BlakeBriddle2005}
{Blake} C., {Bridle} S., 2005, \mnras, 363, 1329

\bibitem[{{Cai} {et~al}\mbox{.}(2009){Cai}, {Angulo}, {Baugh}, {Cole}, {Frenk},
  \& {Jenkins}}]{Cai2009}
{Cai} Y.-C., {Angulo} R.~E., {Baugh} C.~M., {Cole} S., {Frenk} C.~S., {Jenkins}
  A., 2009, \mnras, 395, 1185

\bibitem[{{Chaves-Montero} {et~al}\mbox{.}(2017){Chaves-Montero}, {Bonoli},
  {Salvato}, {Greisel}, {D{\'{\i}}az-Garc{\'{\i}}a}, {L{\'o}pez-Sanjuan},
  {Viironen}, {Fern{\'a}ndez-Soto}, {Povi{\'c}}, {Ascaso}, {Arnalte-Mur},
  {Masegosa}, {Matute}, {M{\'a}rquez}, {Cenarro}, {Abramo}, {Ederoclite},
  {Alfaro}, {Marin-Franch}, {Varela}, \&
  {Cristobal-Hornillos}}]{chavesmontero17}
{Chaves-Montero} J. {et~al.}, 2017, \mnras, 472, 2085

\bibitem[{{Colombi} {et~al}\mbox{.}(2009){Colombi}, {Jaffe}, {Novikov}, \&
  {Pichon}}]{Colombi2009}
{Colombi} S., {Jaffe} A., {Novikov} D., {Pichon} C., 2009, \mnras, 393, 511

\bibitem[{{Crocce} \& {Scoccimarro}(2008)}]{Crocce2008}
{Crocce} M., {Scoccimarro} R., 2008, \prd, 77, 023533

\bibitem[{{Dalton} {et~al}\mbox{.}(2014){Dalton}, {Trager}, {Abrams},
  {Bonifacio}, {L{\'o}pez Aguerri}, {Middleton}, {Benn}, {Dee}, {Say{\`e}de},
  {Lewis}, {Pragt}, {Pico}, {Walton}, {Rey}, {Allende Prieto}, {Pe{\~n}ate},
  {Lhome}, {Ag{\'o}cs}, {Alonso}, {Terrett}, {Brock}, {Gilbert}, {Ridings},
  {Guinouard}, {Verheijen}, {Tosh}, {Rogers}, {Steele}, {Stuik}, {Tromp},
  {Jasko}, {Kragt}, {Lesman}, {Mottram}, {Bates}, {Gribbin}, {Rodriguez},
  {Delgado}, {Martin}, {Cano}, {Navarro}, {Irwin}, {Lewis}, {Gonzalez Solares},
  {O'Mahony}, {Bianco}, {Zurita}, {ter Horst}, {Molinari}, {Lodi}, {Guerra},
  {Vallenari}, \& {Baruffolo}}]{weave14}
{Dalton} G. {et~al.}, 2014, in \procspie, Vol. 9147, Ground-based and Airborne
  Instrumentation for Astronomy V, p. 91470L

\bibitem[{{de Jong}(2011)}]{4most11}
{de Jong} R., 2011, The Messenger, 145, 14

\bibitem[{{DESI Collaboration} {et~al}\mbox{.}(2016){DESI Collaboration},
  {Aghamousa}, {Aguilar}, {Ahlen}, {Alam}, {Allen}, {Allende Prieto}, {Annis},
  {Bailey}, {Balland}, \& et~al.}]{DESI16}
{DESI Collaboration} {et~al.}, 2016, ArXiv e-prints

\bibitem[{{Dolney} {et~al}\mbox{.}(2006){Dolney}, {Jain}, \&
  {Takada}}]{Dolney2006}
{Dolney} D., {Jain} B., {Takada} M., 2006, \mnras, 366, 884

\bibitem[{{Dor{\'e}} {et~al}\mbox{.}(2014){Dor{\'e}}, {Bock}, {Ashby}, {Capak},
  {Cooray}, {de Putter}, {Eifler}, {Flagey}, {Gong}, {Habib}, {Heitmann},
  {Hirata}, {Jeong}, {Katti}, {Korngut}, {Krause}, {Lee}, {Masters},
  {Mauskopf}, {Melnick}, {Mennesson}, {Nguyen}, {{\"O}berg}, {Pullen},
  {Raccanelli}, {Smith}, {Song}, {Tolls}, {Unwin}, {Venumadhav}, {Viero},
  {Werner}, \& {Zemcov}}]{dore14}
{Dor{\'e}} O. {et~al.}, 2014, ArXiv e-prints

\bibitem[{{Dor{\'e}} {et~al}\mbox{.}(2016){Dor{\'e}}, {Werner}, {Ashby},
  {Banerjee}, {Battaglia}, {Bauer}, {Benjamin}, {Bleem}, {Bock}, {Boogert},
  {Bull}, {Capak}, {Chang}, {Chiar}, {Cohen}, {Cooray}, {Crill}, {Cushing}, {de
  Putter}, {Driver}, {Eifler}, {Feng}, {Ferraro}, {Finkbeiner}, {Gaudi},
  {Greene}, {Hillenbrand}, {H{\"o}flich}, {Hsiao}, {Huffenberger}, {Jansen},
  {Jeong}, {Joshi}, {Kim}, {Kim}, {Kirkpatrick}, {Korngut}, {Krause}, {Kriek},
  {Leistedt}, {Li}, {Lisse}, {Mauskopf}, {Mechtley}, {Melnick}, {Mohr},
  {Murphy}, {Neben}, {Neufeld}, {Nguyen}, {Pierpaoli}, {Pyo}, {Rhodes},
  {Sandstrom}, {Schaan}, {Schlaufman}, {Silverman}, {Su}, {Stassun}, {Stevens},
  {Strauss}, {Tielens}, {Tsai}, {Tolls}, {Unwin}, {Viero}, {Windhorst}, \&
  {Zemcov}}]{dore16}
{Dor{\'e}} O. {et~al.}, 2016, ArXiv e-prints

\bibitem[{{Eisenstein} \& {Hu}(1998)}]{Eisenstein1998}
{Eisenstein} D.~J., {Hu} W., 1998, \apj, 496, 605

\bibitem[{{Eisenstein} {et~al}\mbox{.}(2007){Eisenstein}, {Seo}, {Sirko}, \&
  {Spergel}}]{Eisenstein2007}
{Eisenstein} D.~J., {Seo} H.-J., {Sirko} E., {Spergel} D.~N., 2007, \apj, 664,
  675

\bibitem[{{Eisenstein} {et~al}\mbox{.}(2005){Eisenstein}, {Zehavi}, {Hogg},
  {Scoccimarro}, {Blanton}, {Nichol}, {Scranton}, {Seo}, {Tegmark}, {Zheng},
  {Anderson}, {Annis}, {Bahcall}, {Brinkmann}, {Burles}, {Castander},
  {Connolly}, {Csabai}, {Doi}, {Fukugita}, {Frieman}, {Glazebrook}, {Gunn},
  {Hendry}, {Hennessy}, {Ivezi{\'c}}, {Kent}, {Knapp}, {Lin}, {Loh}, {Lupton},
  {Margon}, {McKay}, {Meiksin}, {Munn}, {Pope}, {Richmond}, {Schlegel},
  {Schneider}, {Shimasaku}, {Stoughton}, {Strauss}, {SubbaRao}, {Szalay},
  {Szapudi}, {Tucker}, {Yanny}, \& {York}}]{Eisenstein2005}
{Eisenstein} D.~J. {et~al.}, 2005, \apj, 633, 560

\bibitem[{{Foreman-Mackey} {et~al}\mbox{.}(2013){Foreman-Mackey}, {Hogg},
  {Lang}, \& {Goodman}}]{Foreman-Mackey2013}
{Foreman-Mackey} D., {Hogg} D.~W., {Lang} D., {Goodman} J., 2013, \pasp, 125,
  306

\bibitem[{{Glazebrook} \& {Blake}(2005)}]{GlazebrookBlake2005}
{Glazebrook} K., {Blake} C., 2005, \apj, 631, 1

\bibitem[{{Hartlap} {et~al}\mbox{.}(2007){Hartlap}, {Simon}, \&
  {Schneider}}]{Hartlap2007}
{Hartlap} J., {Simon} P., {Schneider} P., 2007, \aap, 464, 399

\bibitem[{{Hockney} \& {Eastwood}(1981)}]{Hockney81}
{Hockney} R.~W., {Eastwood} J.~W., 1981, {Computer Simulation Using Particles}

\bibitem[{{Howlett} {et~al}\mbox{.}(2015){Howlett}, {Manera}, \&
  {Percival}}]{Howlett2015}
{Howlett} C., {Manera} M., {Percival} W.~J., 2015, Astronomy and Computing, 12,
  109

\bibitem[{{Ilbert} {et~al}\mbox{.}(2009){Ilbert}, {Capak}, {Salvato}, {Aussel},
  {McCracken}, {Sanders}, {Scoville}, {Kartaltepe}, {Arnouts}, {Le Floc'h},
  {Mobasher}, {Taniguchi}, {Lamareille}, {Leauthaud}, {Sasaki}, {Thompson},
  {Zamojski}, {Zamorani}, {Bardelli}, {Bolzonella}, {Bongiorno}, {Brusa},
  {Caputi}, {Carollo}, {Contini}, {Cook}, {Coppa}, {Cucciati}, {de la Torre},
  {de Ravel}, {Franzetti}, {Garilli}, {Hasinger}, {Iovino}, {Kampczyk},
  {Kneib}, {Knobel}, {Kovac}, {Le Borgne}, {Le Brun}, {F{\`e}vre}, {Lilly},
  {Looper}, {Maier}, {Mainieri}, {Mellier}, {Mignoli}, {Murayama}, {Pell{\`o}},
  {Peng}, {P{\'e}rez-Montero}, {Renzini}, {Ricciardelli}, {Schiminovich},
  {Scodeggio}, {Shioya}, {Silverman}, {Surace}, {Tanaka}, {Tasca}, {Tresse},
  {Vergani}, \& {Zucca}}]{Ilbert2009}
{Ilbert} O. {et~al.}, 2009, \apj, 690, 1236

\bibitem[{{Kaiser}(1987)}]{Kaiser1987}
{Kaiser} N., 1987, \mnras, 227, 1

\bibitem[{{Kashlinsky} {et~al}\mbox{.}(2001){Kashlinsky},
  {Hern{\'a}ndez-Monteagudo}, \& {Atrio-Barandela}}]{Kashlinsky01}
{Kashlinsky} A., {Hern{\'a}ndez-Monteagudo} C., {Atrio-Barandela} F., 2001,
  \apjl, 557, L1

\bibitem[{{Koda} {et~al}\mbox{.}(2016){Koda}, {Blake}, {Beutler}, {Kazin}, \&
  {Marin}}]{Koda2015}
{Koda} J., {Blake} C., {Beutler} F., {Kazin} E., {Marin} F., 2016, \mnras, 459,
  2118

\bibitem[{{Laureijs} {et~al}\mbox{.}(2011){Laureijs}, {Amiaux}, {Arduini},
  {Augu{\`e}res}, {Brinchmann}, {Cole}, {Cropper}, {Dabin}, {Duvet}, {Ealet},
  \& et~al.}]{euclid11}
{Laureijs} R. {et~al.}, 2011, ArXiv e-prints

\bibitem[{{Lidz} \& {Taylor}(2016)}]{Lidz2016}
{Lidz} A., {Taylor} J., 2016, \apj, 825, 143

\bibitem[{{Mart{\'{\i}}} {et~al}\mbox{.}(2014){Mart{\'{\i}}}, {Miquel},
  {Castander}, {Gazta{\~n}aga}, {Eriksen}, \& {S{\'a}nchez}}]{pau14}
{Mart{\'{\i}}} P., {Miquel} R., {Castander} F.~J., {Gazta{\~n}aga} E.,
  {Eriksen} M., {S{\'a}nchez} C., 2014, \mnras, 442, 92

\bibitem[{{McQuinn} \& {White}(2013)}]{McQuinn2013}
{McQuinn} M., {White} M., 2013, \mnras, 433, 2857

\bibitem[{Moore(1920)}]{Moore1920}
Moore R.~L., 1920, Bull. Amer. Math. Soc., 26, 412

\bibitem[{{Orsi} \& {Angulo}(2017)}]{orsiangulo2017}
{Orsi} A.~A., {Angulo} R.~E., 2017, ArXiv e-prints

\bibitem[{{Padmanabhan} \& {White}(2008)}]{Padmanabhan08}
{Padmanabhan} N., {White} M., 2008, \prd, 77, 123540

\bibitem[{{Padmanabhan} \& {White}(2009)}]{Padmanabhan2009}
{Padmanabhan} N., {White} M., 2009, \prd, 80, 063508

\bibitem[{{Peacock} \& {Dodds}(1994)}]{Peacock1994}
{Peacock} J.~A., {Dodds} S.~J., 1994, \mnras, 267, 1020

\bibitem[{Penrose(1955)}]{Penrose55}
Penrose R., 1955, Proc. Cambridge Philos. Soc., 51, 406

\bibitem[{{Ross} {et~al}\mbox{.}(2017){Ross}, {Banik}, {Avila}, {Percival},
  {Dodelson}, {Garcia-Bellido}, {Crocce}, {Elvin-Poole}, {Giannantonio},
  {Manera}, \& {Sevilla-Noarbe}}]{Ross17}
{Ross} A.~J. {et~al.}, 2017, ArXiv e-prints

\bibitem[{{Ross} {et~al}\mbox{.}(2015){Ross}, {Percival}, \&
  {Manera}}]{Ross2015}
{Ross} A.~J., {Percival} W.~J., {Manera} M., 2015, \mnras, 451, 1331

\bibitem[{{S{\'a}nchez} {et~al}\mbox{.}(2008){S{\'a}nchez}, {Baugh}, \&
  {Angulo}}]{Sanchez2008}
{S{\'a}nchez} A.~G., {Baugh} C.~M., {Angulo} R.~E., 2008, \mnras, 390, 1470

\bibitem[{{Schmittfull} {et~al}\mbox{.}(2015){Schmittfull}, {Feng}, {Beutler},
  {Sherwin}, \& {Chu}}]{Schmittfull2015}
{Schmittfull} M., {Feng} Y., {Beutler} F., {Sherwin} B., {Chu} M.~Y., 2015,
  \prd, 92, 123522

\bibitem[{{Scoccimarro}(2004)}]{Scoccimarro04}
{Scoccimarro} R., 2004, \prd, 70, 083007

\bibitem[{{Sefusatti} {et~al}\mbox{.}(2016){Sefusatti}, {Crocce},
  {Scoccimarro}, \& {Couchman}}]{Sefusatti16}
{Sefusatti} E., {Crocce} M., {Scoccimarro} R., {Couchman} H.~M.~P., 2016,
  \mnras, 460, 3624

\bibitem[{{Seo} \& {Eisenstein}(2003)}]{SeoEisenstein2003}
{Seo} H.-J., {Eisenstein} D.~J., 2003, \apj, 598, 720

\bibitem[{{Seo} \& {Eisenstein}(2007)}]{SeoEisenstein2007}
{Seo} H.-J., {Eisenstein} D.~J., 2007, \apj, 665, 14

\bibitem[{{Sereno} {et~al}\mbox{.}(2015){Sereno}, {Veropalumbo}, {Marulli},
  {Covone}, {Moscardini}, \& {Cimatti}}]{Sereno2015}
{Sereno} M., {Veropalumbo} A., {Marulli} F., {Covone} G., {Moscardini} L.,
  {Cimatti} A., 2015, \mnras, 449, 4147

\bibitem[{{Smith} {et~al}\mbox{.}(2008){Smith}, {Scoccimarro}, \&
  {Sheth}}]{Smith2008}
{Smith} R.~E., {Scoccimarro} R., {Sheth} R.~K., 2008, \prd, 77, 043525

\bibitem[{{Szapudi} {et~al}\mbox{.}(2001){Szapudi}, {Prunet}, {Pogosyan},
  {Szalay}, \& {Bond}}]{Szapudi01}
{Szapudi} I., {Prunet} S., {Pogosyan} D., {Szalay} A.~S., {Bond} J.~R., 2001,
  \apjl, 548, L115

\bibitem[{{Tassev} {et~al}\mbox{.}(2013){Tassev}, {Zaldarriaga}, \&
  {Eisenstein}}]{Tassev2013}
{Tassev} S., {Zaldarriaga} M., {Eisenstein} D.~J., 2013, \jcap, 6, 036

\bibitem[{{Weinberg} {et~al}\mbox{.}(2013){Weinberg}, {Mortonson},
  {Eisenstein}, {Hirata}, {Riess}, \& {Rozo}}]{Weinberg2013}
{Weinberg} D.~H., {Mortonson} M.~J., {Eisenstein} D.~J., {Hirata} C., {Riess}
  A.~G., {Rozo} E., 2013, \physrep, 530, 87

\bibitem[{{Xu} {et~al}\mbox{.}(2013){Xu}, {Cuesta}, {Padmanabhan},
  {Eisenstein}, \& {McBride}}]{Xu13}
{Xu} X., {Cuesta} A.~J., {Padmanabhan} N., {Eisenstein} D.~J., {McBride} C.~K.,
  2013, \mnras, 431, 2834

\end{thebibliography}



\label{lastpage} \end{document}